\documentstyle[epsf]{article}

\begin{document}

\title{Particle Dark Matter and its Detection\protect\footnote{Updated
        from the Invited Summary Talk given at the {\em NuPECC Workshop
        on Present and Future of the Neutrino Physics in Europe}, Frascati,
        November 1996. An abridged version has been published in the
        ``Neutrino Physics'' chapter of the NuPECC Report ``Nuclear Physics
        in Europe: Highlights and Opportunities'', December 1997. }}

\author{Angel Morales \\
        Laboratory of Nuclear and High Energy Physics \\
        Faculty of Science, University of Zaragoza, Spain\\}

\date{}

\maketitle

\section{Cosmological Dark Matter Models}

There is substantial evidence that most of the matter of the universe is
dark and a compelling motivation to believe that it is mainly of
non-baryonic origin \cite{fab1,pri2,tri3,kol4,fic5,jun6,pri7}. The flat
rotation curves in spiral galaxies indicate that most of their masses is in
form of large Dark Matter (DM) halos of a mass ten times bigger than the
visible mass. Dynamical evidence for the existence of DM at large scales
comes from a variety of observation which leads to the conclusion that the
luminous mass alone cannot explain the dynamics of the celestial bodies.
Gravitational lensing, ROSAT data on X-ray from bremsstrahlung emission of
fast electrons in the hot intergalactic gas provide evidence of DM at these
scales. Clusters of galaxies velocity dispersion implies the existence of DM
in the intercluster space \cite{str8, bah9}. Finally, IRAS data and POTENT analysis on
peculiar velocities of galaxies, COBE results on the Cosmic Microwave
Background Radiation anisotropies, and high red-shift supernova data support
the presence of DM at the largest scales of the universe.

The mean total energy density of the universe, $\overline{\rho }$, is
currently expressed in units of the critical density $\rho _c$, $\Omega
_0\equiv \overline{\rho }/\rho _c$ with $\rho _c=3H_0^2/8\pi $ $G$ ($=h^2$
1.05$\times 10^4$ eV cm$^{-3}$), where $H_0$ is the Hubble parameter $
H_0\equiv $100 h Kms$^{-1}$Mpc$^{-1}$ and G the gravitational constant. $
\Omega _0$ consists of matter density $\Omega _M$, a negligible part of
radiation and, possibly, vacuum energy $\Omega _\Lambda /3H_0^2$. Values of $
h$ range \cite{pri7} between 0.4 and 1 with a long-standing preference for 0.5.
Recent measurements, however, lead to higher values of $h$ \cite{fre10,bir11}.
Measures and estimates of the mean matter density, $\Omega _M$, come from
various sources, with values ranging from 0.01 to about 1. Luminous matter
in galaxies (~10 Kpc) accounts only for 1\% or less. Galactic halos ($50-100$
Kpc) contribute up to a 10\% to the matter density $\Omega _h\sim 0.03-0.1$.
Mass-to-light ratios in clusters ($\sim $~Mpc) \cite{bah9} suggest that
$\Omega _M>0.2$ in agreement with measurements derived from the dynamics of
large scale structures and large scale velocity fields \cite{str8}. At $\sim $~100
Mpc and above, $\Omega _M$ is measured by using the data on peculiar
velocities of galaxies \cite{pee12,vit13,row14} and red-shift surveys based on the
IRAS catalogue. POTENT analysis \cite{ber15,dek16} and IRAS data give a lower limit
of $\Omega _M>0.3$. As the scale of the observed cosmic structures increase,
the value of $\Omega _M$ becomes larger \cite{per17}, approaching the unity value
predicted by inflationary cosmologies and supported by COBE results \cite{smo18}.

Big-bang nucleosynthesis constrains \cite{wal19} the baryon fraction of a critical
universe to be less than a few percent, $\Omega _B.h^2=0.01-0.02$, and so
baryonic dark matter is needed. On the other hand, the large values of
$\Omega$ at increasing scales, together with the smallness of $\Omega _B$
imply that non-baryonic dark matter should be the main component of the
universe. Dark baryons may be in form of black holes, jupiters, white
dwarfs, brown dwarfs, or the generically so-called MACHOS\ (Massive
Astrophysical Compact Heavy Objects). From the cosmological point of view,
two big categories of non-baryonic DM candidates have been proposed: Cold
Dark Matter (CDM) and Hot Dark Matter (HDM), according to whether they were
slow- or fast-moving at the time of the galaxy formation. Their relative
proportion is fixed to properly generate the observed cosmic structures by
gravitational evolution of the scale-invariant primordial density
fluctuations, as CDM produces structures earlier, whereas HDM erases density
fluctuations at small scales \cite{sha20}. The simple CDM \cite{blu21} model
leads to an excess of structure formation at small scales and need to be
mixed \cite{pri7,tur23} with a small fraction of HDM to match the observed
spectral power at all scales \cite{tur23}. That mixed Cold plus Hot Dark
Matter model \cite{pog22} features $\Omega_{CDM}\approx 0.75$,
$\Omega _{HDM}\approx 0.2$, $\Omega _B\approx 0.05$ ($h=0.5$,
$\Omega _\Lambda =0$, $\Omega =\Omega _M=1$). Recent values of the
Hubble constant ($h\approx 0.7$) might favour instead a Cold DM model with a
non-zero cosmological constant $\Omega _\Lambda =0.7$ and $\Omega _M=0.3$
\cite{pri7}.

\section{Dark Matter Candidates}

\subsection{Baryonic Dark Matter}

Baryonic dark matter in form of MACHOS is actively searched for in
on-going surveys of stars in the Large Magellanic
Cloud by looking for possible microlensing effects produced by such objects
when they pass near the line of sight between the observer and the stars
\cite{pac24}. The results of these observations (MACHO \cite{alc25},
EROS \cite{aub26} and OGLE \cite{uda27} experiments) indicate that MACHOS
could account for a significant fraction
of the dark halo. The results of MACHO, for instance after the first two
years of data, conclude that the halo mass in form of MACHOS,
within 50 Kpc, could be as much as 50\% of the total halo mass,
with MACHO masses in the range of 0.05 to 1 solar masses, its most
probable mass being $0.5_{-0.2}^{+0.3}$
$M_{\odot }$. Brown dwarfs might be, then, good candidates to the baryonic
dark matter of the galactic halo. However, the statistics is still too low
(eight microlensing events in all MACHO plus EROS observations) to establish
firm conclusions, as the results depend on the model of both visible and
dark matter of the galaxy.

\subsection{Non-Baryonic Dark Matter}

Extensions of the Standard Model of Particle Physics provide non-baryonic
dark matter candidates \cite{pri2,jun6,ber28}. They should have non-zero masses, be
electrically neutrals and interact weakly with ordinary matter. On the other
hand, their present relic abundance must have the right values to fill the
gap between $\Omega _B$ and $\Omega _M$.

\subsection{Neutrinos as hot Dark Matter Candidates}

Archetypical hot relics are the ``standard'' light (m$<$MeV) neutrinos \cite{kol4}
\cite{dol29}. Their number density, for each flavor $\nu _e$, $\nu _\mu $, $\nu
_\tau $ is $n_\nu \sim 100$ cm$^{-3}$ and, if endowed with a non-zero mass,
they could provide by themselves the whole critical density. In fact, the
Gerstein-Zeldovich bound $\sum m_\nu \approx \Omega _\nu h^2$92 eV (where
the sum extends over all species of light neutrinos with full weak
interaction) would lead to the right relic density for various neutrino
masses according to the fraction of hot dark matter contained in the model.

In the old, classical hot DM model, neutrinos (one or more species) of
masses $20\sim 30$ eV were supposed to constitute the whole DM, but being
relativistic particles at freeze out, they cannot form galaxies (the
so-called up-down scenario of galaxy formation where larger structures form
earlier). Although cosmic strings would help, cold dark matter is required
\cite{tur23}. In the mixed CHDM model, where the hot DM is $\Omega _\nu =0.2$, the
preferred total mass of the dark matter neutrino would be $\sim 5$ eV\ (for
$h=0.5$) \cite{pri7,pog22} shared between the various neutrino species, according to
the mass pattern used to solve, simultaneously, other neutrino puzzle (for
instance neutrino oscillations). There is no known method proposed so far to
directly detect the hot DM relic neutrinos and so terrestrial sources are
used to explore this possibility. The discovery of a $\nu _\tau $ mass in
the few eV range would favor this form of DM and so several oscillation
experiments are under way to explore that range \cite{dum30}.

\subsection{Cold Dark Matter: Weak Interacting Massive Particles (WIMPs)}

In the CDM sector, typical candidates are heavy Dirac or Majorana neutrinos
in the GeV-TeV mass range or other heavy, weakly interacting neutral
particles, generically called WIMPs \cite{pri2,jun6,tur23,lee31,ell32,hab33}.
WIMPs could have
been produced in the early universe and decoupled at a density given by
their annihilation cross-section through
$\Omega _\chi h^2\sim 10^{-37}\ \rm{cm}^2/\sigma _{annh}$, and so be
cosmologically interesting for interactions of
weak strength. A distinguished Majorana WIMP is the neutralino \cite{ell32,hab33,
gri34,sre35,giu36,dre37,jun6}, the lightest supersymmetric particle of SUSY
theories, which is stable if R-parity $R=(-1)^{3B+L+2S}$ is conserved. In
the minimal supersymmetric standard model (MSSM) \cite{hab33}, the neutralino is a
mixture of the supersymmetric partners of the hypercharge gauge boson B, of
the neutral SU(2) gauge boson $W_3$, and of the neutral components of the
Higgs doublets $H_1^0$, $H_2^0$. The mixture is written as a combination of
photino, zino and higgsinos, $\chi =a_1\widetilde{\gamma }+a_2\widetilde{Z}
+a_3\widetilde{H}_1^0+a_4\widetilde{H}_2^0$. The parameters of this
combination are the higgsinos mixing mass parameter m, the gauginos masses
$M_1$, $M_2$ (of $\widetilde{B}$, $\widetilde{W}_3$ respectively) and
$tg\beta =v_u/v_d$ where $v_u$ and $v_d$ are the vacuum expectation values of
$\widetilde{H}_1^0$, $\widetilde{H}_0^{}$ giving masses respectively to
quarks of type up and down. When implemented with grand unification, only
three parameters are left ($M_1\sim M_2/2$). Accelerator results \cite{ell38}
constrain the neutralino mass, for representative values of its parameter
space, to be above 10--20 GeV and conclude that it is rather a whole mixture
of the four components, not a simple one---like pure photino or pure
higgsino. There exist however unconstrained MSSM models \cite{gab39} where
neutralinos of masses as smaller as 2 GeV may exist. Neutralino relic
abundances of cosmological interest are in the range $\Omega _\chi h^2\sim
0.2$ \cite{ber28,bot40,bot41,gri42,gri34} and are found for a wide range of
values of the neutralino parameter space. The choice of these parameters
will also determine the detectability of the neutralino \cite{jun6,bot40,bot41,ber76}.

\subsection{Axions and their detection}

Another celebrated CDM candidate is the axion \cite{pec43}, a non-thermal relic
invented to solve the strong CP problem. The axion, a light neutral pseudo
scalar Goldstone particle of very weak interaction, emerging from the
spontaneous breaking of the Peccei-Quinn symmetry, could provide the
universe critical density \cite{kol4,tur44,raf45} for masses of about $10^{-5}$ eV and
number density of $10^9$ cm$^{-3}$.

Galactic axions can be converted into photons in a strong magnetic field
\cite{sik46}, feature which has inspired the axions searches \cite{hag47,wue48,bib49}.
The search for relic axions through the Primakoff effect is performed with
superconducting resonant cavities \cite{hag47,wue48,bib49,oga50}, and recently with
CERN's SMC polarized target \cite{sem51}. The signature for the axion is a narrow
peak at frequency $v$, resulting from its resonant conversion into a photon
at $hv=m_ac^2$. The sensitivity of these experiments is limited by the
possibility that the axion rest mass lays outside resonance for the employed
cavities. Pioneering experiments were carried out at the University of
Florida \cite{wue48} and Brookhaven \cite{hag47}.
A collaboration MIT/LLNL/Florida/FNAL/UC
Berkeley/INR Moscow is running a large scale RF cavity halo axions detector,
taking data since the beginning of 1996 \cite{hag52}. The superconducting magnet is
a six-tons niobium-titanium solenoid (110 cm long, 60 cm $\emptyset $),
producing a 8.5 Tesla field. The first data have scanned the 2.8--3.3 $\mu $eV
mass regions. The parameter space to be explored by this experiment at its
current sensitivity is between 1.3 and 13 $\mu $eV. Further improvements (a
factor 10 in sensitivity) are foreseen by using ultralow noise DC SQUID
amplifiers.

Energetic axions can be produced continuously in the interior of stars (red
giants, supernovae), or in the Sun . Astrophysical (cooling rates of stars)
and cosmological (overclosure of the universe) arguments, as well as
laboratory experiments, require an axion mass in the range $10^{-6}$ eV
$<m_a<10^{-3}$ eV \cite{raf45,sik46,che53,raf54,kra55,jan56}. Other masses are
possible for hadronic axions, not coupled directly to leptons and interacting
with matter through a two-photon vertex. The red giants bound stands at
$g_{a\gamma \gamma }<10^{-10}$\thinspace GeV$^{-1}$ \cite{raf45}.

The Sun is a powerful source of axions in the $1\sim 15$ keV energy range.
The solar axion telescope experiment of the Rochester-BNL-FNAL collaboration
\cite{laz57} got the coupling limits:
$g_{a\gamma \gamma }<3.6\times 10^{-9}\ \rm{GeV}^{-1}$ for
$m_a<30$ meV and $g_{a\gamma \gamma }<7.7\times 10^{-9}\ \rm{GeV}^{-1}$
for $30<m_a<110$ meV. A method of detecting solar axions was
proposed in Ref. \cite{buc58,pas59} and extended recently in Ref. \cite{cre60}.
The axions convert coherently into photons in the lattice of a germanium crystal
when the incident angle satisfies the Bragg condition. As shown in \cite{cre60}, the
detection rates in various energy windows are correlated with the relative
orientations of the detector and the sun. This correlation results in a
temporal pattern which should be a distinctive, unique signature of the
axion. A recent Ge detector experiment \cite{avi61} has provided a new laboratory
bound of $g_{a\gamma \gamma }<2.7\times 10^{-9}$ GeV$^{-1}$, independent of
axion mass up to $\sim $1 keV.

\section{Searches for Non-baryonic Dark Matter:\ General Features of the
Detection. Strategies and Techniques \cite{jun6,pri2,smi62}}

Discovering the nature of the dark matter is one of the big challenges in
Cosmology, Astrophysics and Particle Physics. There exists a large activity
going on in non-baryonic dark matter searches through indirect or direct
detection methods. No DM signal has been detected so far, but various kinds
of candidates have already been excluded or constrained.

\subsection{WIMPs Indirect detection. Large Underground Detectors and
Neutrino Telescopes}

Particle Dark Matter can be detected indirectly by searching in cosmic ray
experiments for particles produced in the WIMP annihilation in the galactic
halos \cite{sil63,sal64}, like antiprotons, positrons or photons. The upcoming
projects PAMELA \cite{ahl65} and AMS \cite{adr66} will provide information on
the antiproton component in cosmic rays and its implication on WIMPs annihilation
in the halo. Dark Matter can also be detected by looking for the high energy
neutrinos emerging as final products of WIMPs annihilation in celestial
bodies \cite{gou67}, Sun \cite{sil68,kra69} or Earth \cite{kra70,fre71,gai72},
in deep underground detectors, or in running underwater (or
underice) neutrino telescopes \cite{bot73,hal74,ber75,ber76,hal77,bot41}.

WIMPs orbiting through the Earth or the Sun can be trapped inside these
bodies when their velocity, as a result of a series of scatterings with
the Earth or the Sun nuclei, drops below the escape velocity from the celestial
body. The WIMPs sink gradually to the center where they accumulated and eventually
annihilate each other into leptons, quarks, $Z^0$, $W^{\pm }$, Higgs...,
which finally give rise to high energy muon neutrinos. By interacting with
the surrounding medium of a detector (neutrino telescope), such neutrinos
produce muons, which are the indirect signature of WIMPs.

Simple kinematics show that the average energy of the emitted neutrino is
about 30\% to 50\% of the WIMP mass, i.e. significantly larger than that of
other neutrinos coming from the Sun or Earth, a fact which permits
distinguish the WIMP neutrinos \cite{hal77}. Further subtraction of the atmospheric
neutrino background leaves the way open for a most promising method of
indirect WIMP identification, zeroing into the zone of GeV--TeV the search
for neutrinos coming from the Sun or the Earth. This capability of the
telescopes will be probably unmatched by the direct detection methods in the
case of high mass WIMPs and spin-dependent couplings.

\begin{figure}[bt]
\centerline{
\epsfxsize=6cm
\epsffile{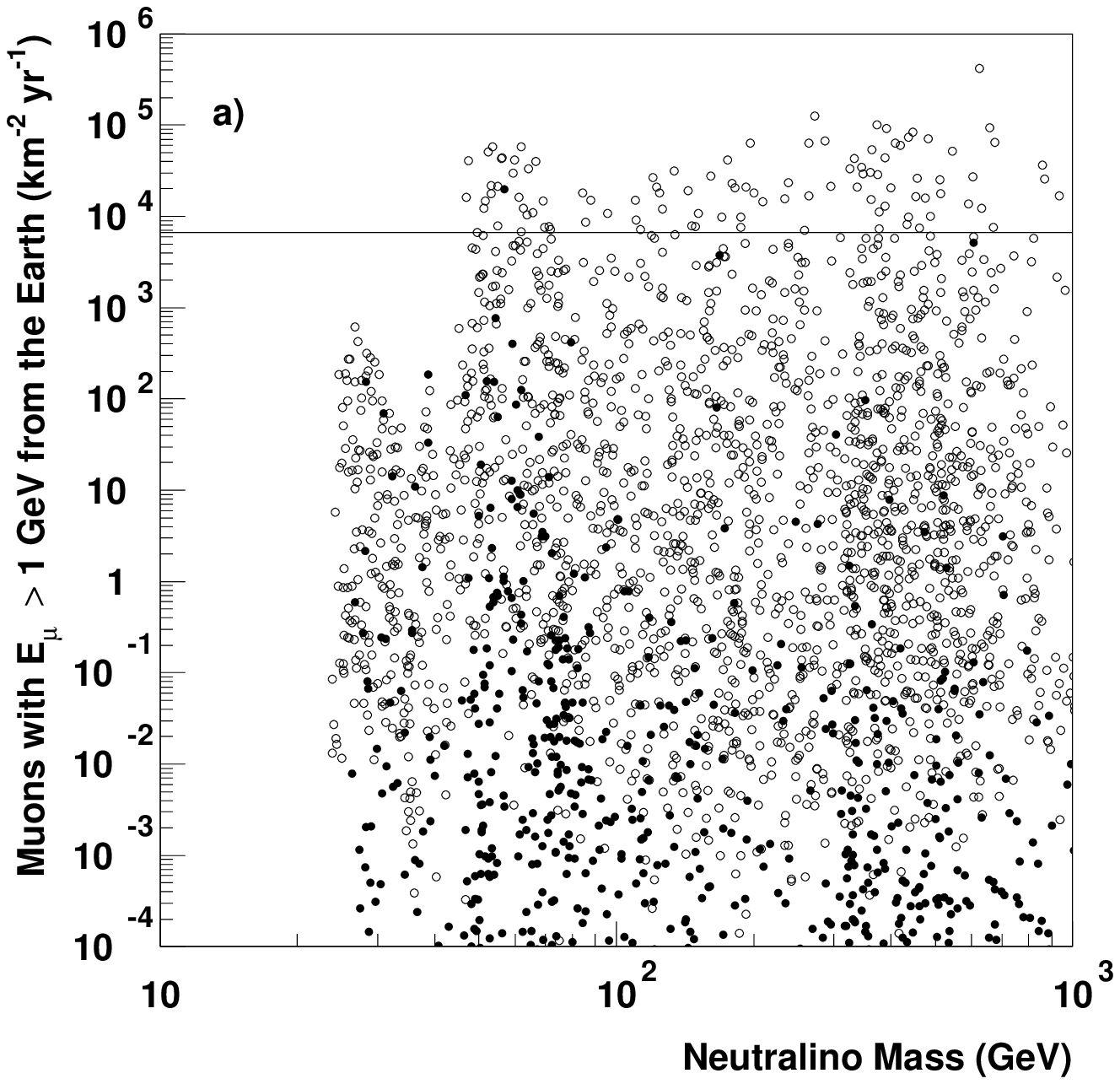}
\epsfxsize=6cm
\epsffile{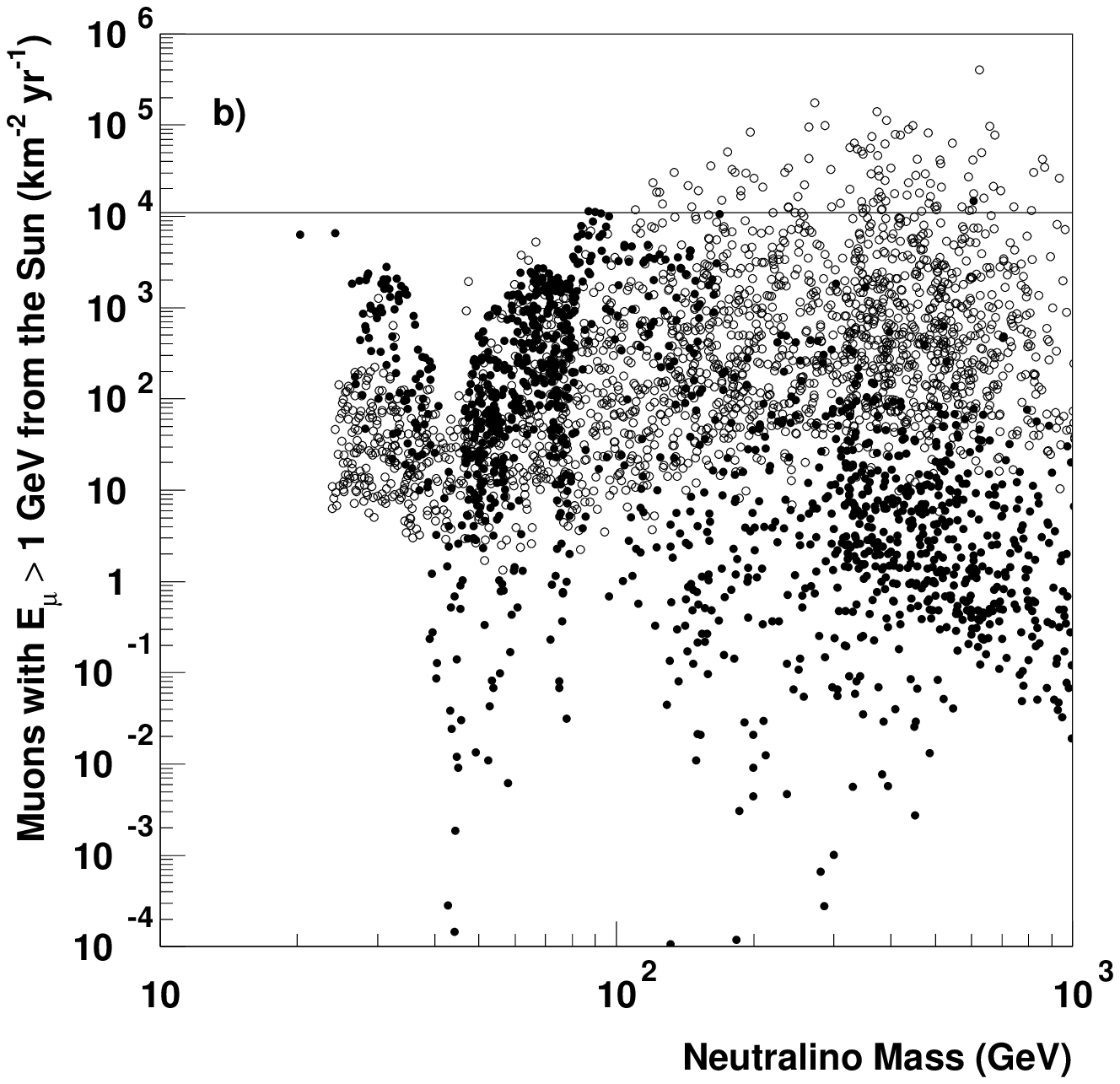}
   }
\caption{Predicted muon fluxes (from WIMPs annihilation) from the Earth
(a) and from the Sun (b), Ref. \cite{ber76} compared with the Baksan limit
\cite{bol79}}
\label{fig1}
\end{figure}

Multipurpose large underground detectors (MACRO \cite{mon78}, Frejus,
Baksan \cite{bol79}, Soudan), and Neutrino Cerenkov telescopes (IMB \cite{los80},
Kamiokande \cite{mor81}), have been used to look for WIMPs neutrinos.
We mention here the constrains to DM particle parameters obtained from
Kamiokande, Baksan and MACRO. MACRO (Monopole Astrophysics and Cosmic Ray
Observatory) in Gran Sasso has searched \cite{mon78} for neutrino-induced
upward-going muons
coming from the direction of the Sun or of the Earth core collected with the
lower part of the detector. No statistical significant signal has been
discriminated over possible fluctuations of the atmospheric neutrinos
background. The muon flux limit from non-atmospheric origin (in the
$25^{\circ }$ window) is $3.1\times 10^{-14}$ cm$^{-2}$ s$^{-1}$ (from the
Earth) and of $6.6\times 10^{-14}$ cm$^{-2}$ s$^{-1}$ (from the Sun), after
exposures of 1250 m$^2$ yr and 380 m$^2$ yr respectively. Indirect searches
of neutralinos have been carried out at Baksan \cite{bol79} (at 850 m.w.e.) with the
scintillator telescope ($17\times 17\times 11$ m$^3$). In 11.94 years of
data, the upper bound on the muon fluxes produced by neutrinos of
non-atmospheric origin are $2.1\times 10^{-14}$ cm$^{-2}$ s$^{-1}$ (90\%
C.L.) and $3.5\times 10^{-14}$ cm$^{-2}$ s$^{-1}$ (90\% C.L.) (from the
Earth and the Sun respectively), corresponding to exposures of 2954 m$^2$ yr
and 1002 m$^2$ yr. These results improve those of Kamiokande \cite{mor81},
$4.1\times 10^{-14}$ cm$^{-2}$ s$^{-1}$ (Earth) and
$6.6\times 10^{-14}\ \rm{cm}^{-2}\ \rm{s}^{-1}$ (Sun) obtained with
exposures of 770 m$^2$ yr and 215 m$^2$ yr.
Predictions for the fluxes of such muons in supersymmetric models
range in the case of the Earth from about $10^{-14}$ cm$^{-2}$ s$^{-1}$ to
$10^{-17}$ cm$^{-2}$ s$^{-1}$ (and somehow higher for the Sun case), and so
bounds on the neutralino annihilation rate as a function of the neutralino
mass have been obtained and regions of the neutralino parameter space
excluded \cite{ber75,ber76,bot41} as depicted in Figure \ref{fig1} \cite{ber76},
corresponding to the scattered plots of the predicted WIMP muons coming from
the Earth (Fig. 1a) and from the Sun (Fig. 1b), compared with the Baksan
limit \cite{bol79}. The above mentioned underwater neutrino telescopes project to
enlarge the exposure areas to reach the $10^{-15}$ cm$^{-2}$ s$^{-1}$ flux
limit. Surfaces larger than $10^5$ m$^2$ would be suitable devices
\cite{jun6,bot73,hal74,ber75,ber76,hal77,bar83} to search for neutralinos in
wide zones of their parameter space and so projects for larger deep
underwater neutrino telescopes (1Km$^3$) are underway \cite{nut87}.

The main interest in the large neutrino telescopes \cite{hal77,res84,lea85,mos86}
is however motivated by their capability as high energy neutrino explorers of
galactic and extragalactic sources \cite{hal77,lea85,mos86}. They are expected
to provide unique information about the astrophysical sources of cosmic rays of
energies as high as $5\times 10^{20}$ eV. The cosmic sources or ``cosmic
accelerators'' which originate ultrahigh energy gamma rays (Active Galactic
Nuclei, pulsars, X-ray binary systems, remnants of young supernovae) are
supposed to be also copious sources of neutrinos through various mechanisms.
Such neutrinos can be detected through the muons they produce in charge
current interactions with the matter surrounding the detector. The higher
the neutrino energy, the smaller the deflection of the outgoing muon from
the neutrino incident direction, and so the emerging muon points back to the
neutrino source. Neutrino telescopes will produce a ``neutrino sky map'' to
complement the current sky image we have from other carriers, techniques and
instruments (infrared, ultraviolet, macro and micro radio wave, CMBR, high
and very high energy gamma rays...). They will allow also to explore regions
not reachable through gamma observations due to the interstellar matter or
to its absorption by the microwave cosmic background. Besides the indirect
detection of WIMPs, neutrino telescopes could address
also the exploration of other big-bang relics, cosmic strings, topological
defects, which might emit, when collapsing, particles with energies up to
$10^{25}$ eV, and the subsequent production of neutrinos. Needless to mention
the usefulness of the telescope in the study of atmospheric neutrino
oscillation or in long-baseline neutrino oscillation experiments.

Various estimates concluded that Neutrino Astronomy requires a mega-telescope of
$\sim $1Km$^3$ \cite{nut87,hal77}, which should be obviously placed deep
underwater of under ice. A neutrino water Cerenkov telescope is a
tridimensional array of optical modules (pressure vessel containing standard
PM tubes and associated electronics) suspended on strings, which detects the
Cerenkov light produced by the neutrino-induced muons. The measured light
intensity---proportional to the muon energy---provides an upper limit to the
neutrino energy. The muon trajectory is reconstructed from the time arrival of
the Cerenkov wave front to the various PMs hit. Corresponding time resolution
of the PMs are a few nanoseconds. Angular resolutions of about one degree can
be obtained [$\theta _{\mu \nu }<1.5^{\circ }\ \sqrt{E}$(TeV)]. The position of
the tridimensional set of optical sensors should be accurately monitored.
Besides the technical challenges to be solved in the telescope construction,
there exist many basic issues to be addressed, in particular, the
determination of the physical parameters of the detector medium as well as
the optimization of the geometry and instrumentation of the PM array for a
given scientific objective. The net spacing, instrumentation and fiducial
volume of the neutrino telescope can be tailored according to the specific
purpose of the instrument. A first stage (0.01 Km$^2$ of surface) of a modular,
multipurpose telescope would be a very interesting tool for atmospheric
neutrino experiments. Surface dimensions of 0.1 km$^2$, with a high energy
threshold could be already useful for neutrino astronomy. A device of
similar dimensions, with lower threshold and denser instrumentation would
perform as WIMP indirect detector.

The pioneering contribution of DUMAND (Deep Underwater Muon and Neutrino
Detector) \cite{bos88}, now abandoned, has settled the frame where the current
telescopes or projects have been developed. Substantial progress in such
projects has been accomplished in the past few years \cite{nut87}.

The AMANDA \cite{hul89,nut87} (Antarctic Muon and Neutrino Detector Array) detector
deployed first four strings of optical modules at a depth from 810 to 1000
meters (Phase A) in the Antarctic polar ice, and then added seven more
strings between 1520 to 1900 m deep (Phase B). A total of 290 optical
sensors have been operating. Now AMANDA-II, featuring ten more strings around
the A+B setup is in progress toward an effective exposure area of 60.000
m$^2 $, and angular opening of one degree. Scattering lengths of Cerenkov
light in AMANDA B are two orders of magnitude larger than in AMANDA A. The
atmospheric muon flux has been measured to be 25 Hz in AMANDA B. Coincidence
measurements between the sectors A and B have proved that up- and
down-going events are well distinguished.

ANTARES \cite{mon90,nut87} (Astronomy with a Neutrino Telescope and Abyss Research)
plans to place in a 2600 m deep abyss off-shore of the Toulon-Marseille
Mediterranean coast a ``demonstrator'' consisting of three strings of 32
optical sensors each. In the future this instrument could be enlarged to the
cubic kilometer scale with 49 strings (100 to 150 meters apart) and a total
of 1500 PM. Supporting mechanical structures have been successfully deployed
in a harbor basis and also at sea (2300 meters deep), 30 Km off shore.
Optical background due to bioluminisicence and to $^{40}$K in water has been
measured. A first step will be the development of 30 Benthos spheres (with
various instrumented optical modules) connected by electro-optical cables to
the shore station.

The Baykal \cite{bez82,spi91} lake neutrino telescope consists of a set of strings
pending from an umbrella-like structure, instrumented with PM in large
vessels, which has been operating since 1993 at a depth of 1366, 3.6 km
off-shore. After the first instrument NT-36 (36 modules in three strings),
which operated along 300 days in 1993-1994 and was successfully recovered, a
larger set (NT-96) has been operating along 1996-1997. After analyzing
$6.5\times 10^7$ events in NT-36, two clear neutrino induced upward going
muons events were detected (vs 1.2 expected from atmospheric neutrinos) and
another three in NT-96. At the time of this writing, (fall of 1997), they
have 144 modules and the NT-200 instrument is supposed to be completed by
mid 1998.

NESTOR \cite{res84,mon92} (Neutrinos from Supernova and TeV sources Ocean Range),
will be installed at 3800 meters depth off-shore from Pylos (Greece).
Successful deployments and recovery at 2600 meters deep of two mechanical
structures of six articulated arms have been accomplished. The structure
forms a star in aluminum (and in titanium), and the optical modules are
pending at the end of each arm of the star. Each star constitutes one
hexagonal floor (of 16 meters of diameter) in a projected tower of 12
floors, containing 168 PMTs per tower. A total of seven towers (1176 optical
modules) is planned. Technical work on the deployment at deep sea, as well
as measurement and characterization of the water physical properties have
been accomplished.

The four collaborations are engaged in active R+D projects aiming to a
future 1Km$^3$ neutrino telescope. The construction of the Km$^3$ scale
instrument would likely proceed in several steps---scientifically
interesting by themselves---to solve gradually the formidable technical
tasks in its development and at the same time perform, at each step,
relevant research on neutrino physics. A sequence of dimensions 0.01 Km$^3$
--0.1 Km$^3$--1Km$^3$ has been suggested \cite{nut87} corresponding to a significant
research on neutrino oscillation, WIMP indirect searches and high energy
neutrino astronomy respectively.

\subsection{WIMP direct detection}

\subsubsection{General features.}

Particle Dark Matter can also be detected through its direct interaction
with ordinary baryonic dark matter \cite{pri2,smi62,goo93,was94}.
The non-relativistic ($\sim $0.001 c) and heavy (TeV--GeV) dark matter
WIMPs supposedly forming the galactic haloes could make a nucleus recoil
with a few keV, from which only a fraction is visible in the detector.
The detection rate depends on the halo model, on the type of WIMP and
interaction, on the nuclear target and detector, as well as on the nuclear
recoil relative efficiency in producing ionization, scintillation or heat
(quenching factor). Semiconductor detectors of Ge and Si, scintillators---both
solid (NaI, CaF$_2$, LiF) and liquid (Xe)---and thermal (bolometer) detectors
of Si, Ge and sapphire, have been used so far \cite{ber95,avi96,cal97}.

The typical rates that would be produced, for instance, by WIMPs of Dirac
neutrino-like type having spin-independent nuclear coherent interaction \cite{goo93}
with Ge nuclei through $Z^0$ exchange reach a few hundred counts per
Ge-detector kilogram and day, and so could be larger than the background
achieved in the best ultralow background Ge detectors (0.1 to 2 counts per
keV kg day), feature which has been used to exclude \cite{cal104} that type of
candidates for a wide region of masses in various underground Ge experiments
\cite{ahl98,dru99,cal100,reu101,mor102,bec103}. The rate that would be due to
axial, spin dependent interactions \cite{ell123} (like that of Majorana neutrino-like
particles) with detector nuclei having non-zero spin would be much smaller, and
so more difficult to attain experimentally. Nevertheless, valuable exclusions for
such spin-dependent couplings have been obtained in NaI
\cite{bac105,dav106,sar107,fus108}, CaF$_2$ \cite{bac109,haz132} and Xe
\cite{bel110} underground experiments.

More appealing candidates, such as the neutralino are starting to be at hand
with the new generation of detectors, in particular in the case of the
coherent neutralino nucleus spin-independent interaction through Higgs
exchange, which might go up to a few counts per kg and day for some regions
of their parameter space \cite{bar83,bot40,bot41,jun6}. Some regions have already
been excluded using Ge and NaI detectors data \cite{bot40,bot127}. The spin-dependent
contribution in the neutralino-nucleus scattering---through $Z^0$ or squark
exchange---is still below the current background achieved.

The main features of the DM interaction---small energy deposits and small
rates---settle the strategy of the DM direct searches: To use detectors of
very low energy threshold and very low background (intrinsic and
environmental), in particular in the low energy region where the nuclear
recoil produced by the WIMP\ scattering is expected. The recorded
background, sets the level of exclusion of the particle dark matter
candidate and it must be reduced at its minimum level by employing passive
and active shielding in a clean environment and using selected material of
very low intrinsic radioactivity. An obvious requirement is to perform the
experiment in an underground site.

A further step to reduce the background is the use of mechanisms
distinguishing those events due to electron recoils (tracers of the
background) from those due to nuclear recoils. Hybrid detectors measuring at
the same time the ionization and heat (or the light and heat) produced in
the detector, or the use of pulse shape discrimination techniques are
successful attempts in the quest for the background reduction.

The possible presence of a particle CDM component in a spectrum is obtained
by comparing the predicted CDM signal with the experimental spectrum. If the
predicted WIMP-nucleus interaction rate is larger than the number of
observed counts in a given energy region, the particle under consideration
can be ruled out as a dark matter component for the values of masses and
cross-sections---or for the corresponding regions of the parameter space
defining the particle dark matter candidate---for which such rate was
derived. The results are usually expressed as an exclusion plot in the plane
of the WIMP-nucleus elastic scattering cross section, $\sigma _{\chi N}$
versus the WIMP mass $m_\chi $.

This conventional method of simply comparing the expected signal with the
observed background spectrum (which may mimic itself the signal) is not
supposed to detect the tiny imprint left by the dark matter particle, but
only to exclude or constrain it. A convincing proof of the detection of CDM
would be to find distinctive signatures in the data characteristic of the
CDM, not faked by the background or by instrumental artifacts. Various DM
identification signatures have been proposed: an annual modulation
\cite{dru111} due to the seasonal June-December variation in the relative
velocity Earth-halo; a very tiny daily modulation \cite{col112} due to the
Earth progressive eclipsing along the day of the DM halo particles in its
way to the detector; a forward-backward asymmetry \cite{spe113} in the direction
of the nuclear recoil due to the Earth motion through the halo, and the
nuclear target dependence of the rate \cite{smi62}. The search for annual
modulation is by now the most extended and promising strategy of DM identification.

On the other hand, only a small fraction of the energy delivered by the WIMP
goes to ionization, the main part being released as heat. More efficient
excitation mechanisms, like Cooper pairs breaking in superconductors or
phonons in thermal detectors (involving typical excitation energies of meV
or $\mu $eV respectively), should be employed. Such cryogenic detectors
\cite{boo114,cry115,sup116,fio117,sad152} will achieve lower energies
threshold, almost 100\% efficiency and have better energy resolution
than the conventional detectors.

\subsubsection{Detection rates.}

WIMPs interact with quarks within nucleons inside a nucleus, and
consequently the WIMP-nucleus cross-section depends on properties and
parameters encompassing such three levels. In the non-relativistic
approximation, the WIMP-nucleus cross-section has two components: a
spin-independent part, and an axial, spin-dependent part, corresponding
respectively to the coupling of WIMPs to the mass or to the spin of the
nucleus \cite{goo93,jun6}.

The differential recoil energy spectrum produced by WIMPs interacting with
nuclei is calculated straightforwardly \cite{goo93,was94,jun6,lew118} once
the halo model and the type of WIMP and interaction have been defined. The
spectrum should be properly corrected for the nuclear recoil relative efficiency
in producing ionization, scintillation or heat, as well as for the loss of
coherence (for momentum transfers larger than the inverse nucleus radius).
The nuclear physics for the spin-independent case is well approximated by a
simple form factor \cite{ahl98,eng119}. The axial case, however, requires detailed
nuclear models \cite{eng120,res121}. The different values chosen for the parameters
entering in the various levels of the WIMP-nucleus interaction, as well as
the large parameter space defining the WIMP (say the relic neutralino), make
the theoretical rate encompass various orders of magnitude.

The interaction rate of WIMPs producing a recoil $T$ in the detector is
given by

\begin{equation}
\label{eq1}\frac{dN}{dt\ dT}=N_D \frac{\rho_h}{m}\int_{v_{min}(T)}^{v_{max}}
\frac{d\sigma}{dT} (v,T) f(\vec {v})vd^3v
\end{equation}

where $T$ is the recoil energy of the target nucleus, $f(\overrightarrow{v})$
is the velocity distribution of the halo DM particle in the reference frame
of the Earth, $N_D$ is the total number of target nuclei, say per kilogram,
$\rho _h$ is the local galactic halo density, and $d\sigma /dT$ is the
WIMP-nucleus differential scattering cross section. The mass of the WIMP
particle is denoted by $m$. The measured deposited (electron equivalent or
visible) energy $E$ and the nuclear recoil energy $T$ are related by a
relative efficiency factor $Q$ (or quenching) $E=TQ$ (nuclear versus
electron recoil efficiency in producing ionization). The quantity
$\overrightarrow{v}_{\min }(T)$ is the minimum relative velocity a particle
must have in order to leave an energy $T$ in the detector, i.e.
$\overrightarrow{v}_{\min }^2=T(M+m)^2/2m^2M$, where $M$ is the target
nucleus mass, and $\overrightarrow{v}_{\max }$ is the maximum velocity of
WIMPs relative to the detector, i.e., the vectorial sum of the galactic
escape velocity $\overrightarrow{v}_{esc}$ and the velocity of the Earth
through the halo, $\overrightarrow{v}_r$. It is customarily assumed that the
DM forms a non-rotating, isothermal, and spherically symmetric halo. In the
galactic rest frame, the WIMPs are supposed to have a Maxwellian velocity
distribution, with a velocity dispersion $v_{rms}$. The local halo density
is customary taken $\rho =0.3$ GeV cm$^{-3}$ with the caution that this
value could have an uncertainty of a factor 2 \cite{cal122}. In fact, a flattened
halo distribution leads to $\rho =0.51_{-0.17}^{+0.21}$ GeV cm$^{-3}$ from
microlensing data.

The WIMP interaction rate in counts per keV per kilogram and day reads

\begin{equation}
\label{eq2}S=7.76 \times 10^{14}N_D \frac{1}{Q} \frac{(m+M)^2}{4Mm^3}
\sigma_W\ \rho \frac{\eta}{v_r}
\end{equation}

where $\sigma _W$ is the elastic cross section of the WIMP-nucleus
interaction (in cm$^2$), which includes the form factor correction for high
momentum transfer, $\sigma _W=\sigma F^2(T)$, $\sigma $ being the point-like
nuclear cross-section. The integral over the velocity distribution appears
through the function $\eta $ given in the simple case of an infinite escape
velocity by $\eta =erf(x+y)-erf(x-y)$ where $x=\sqrt{3/2}v_r/v_{rms}$ and
$y=\sqrt{3/2}v_r/v_{rms}$ and $erf(z)$ the usual error function $\left(
erf(z)=2/\sqrt{\pi }\int_0^z\exp (-t^2)dt\right) $.

The cross-section for the spin-independent heavy neutrino (or
neutralino)-nucleus interaction, is strongly enhanced by the coherence
factor $\sigma ^{SI}\sim N^2(A^2)$. In the case of Majorana WIMPs (like
heavy Majorana neutrinos, neutralinos), the spin-dependent cross-section is
proportional to the nuclear spin, $\sigma ^{SD}\sim \Lambda ^2J(J+1)$
\cite{eng120,res121,ell123}, where
$\Lambda =(a_p\left\langle \overline{S}_p\right\rangle
+a_n\left\langle \overline{S}_n\right\rangle )/J$, $\left\langle \overline{S}
_{p,n}\right\rangle $ are the contributions of the total proton and neutron
spin in the nucleus and $a_{p,n}$ the coupling constants WIMP-nucleon,
depending upon the type of WIMP and interaction and on the quark spin
distribution within the nucleon. Typical integrated rates for Ge are $\sim
10^2$ c/kg day, for the heavy Dirac neutrino spin-independent contribution;
$\leq $1 c/kg day for the coherent contribution neutralino-nucleus
cross-section (due to Higgs exchange); and $\leq $0.01 c/kg day for the
spin-dependent contribution to the $\chi $-nucleus scattering (due to $Z^0$
exchange) for typical values of the parameters space \cite{bot40,bot41}.

\subsubsection{WIMPs searches with Ge ionization detectors.}

In the first generation of experiments, Si and Ge diodes were used. Various
experimental groups have searched for WIMPs with Ge detectors as byproducts
of double beta decay experiments: USC/PNL in Homestake \cite{ahl98,dru99};
USCB / UCB/ LBL in Oroville \cite{cal100}; Caltech/PSI/Neuchatel in Gothard
\cite{reu101}; Zaragoza/ USC/PNL in Canfranc \cite{mor102}; and Heidelberg/Moscow
in Gran Sasso \cite{bec103}, with energy thresholds ranging from 1.6 keV to 12
keV (electron equivalent energy), and backgrounds at threshold of
0.2 to 5 c/keV.kg.day (lower backgrounds are obtained at a few keV
above thresholds).

\begin{figure}[bt]
\centerline{
\epsfxsize=8cm
\epsffile{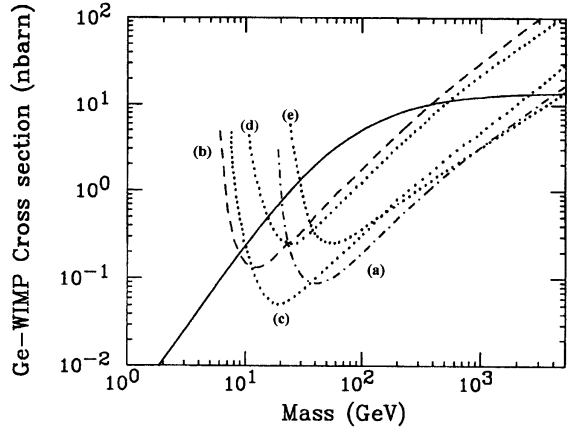}
   }
 \caption{Exclusion plots for WIMPs spin-independent coupling (Ge):
 (a) Ref. \cite{dru99}; (b) Ref. \cite{mor102}; (c) Ref. \cite{reu101};
 (d) Ref. \cite{cal100}; (e) Ref. \cite{bec103}}
  \label{fig2}
\end{figure}

\begin{figure}[bt]
\centerline{
\epsfxsize=8cm
\epsffile{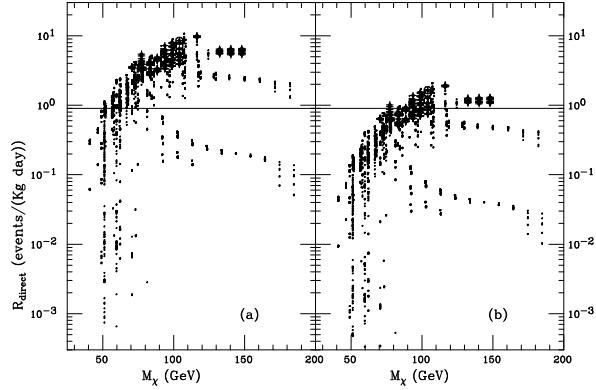}
   }
 \caption{ Scatter plots of the rate (neutralino/Ge) as a function of the
 neutralino mass, for various neutralino configuration of different cosmological
relevance \cite{bot40}. The horizontal line is the Homestake (Germanium
experiment) limit \cite{dru99}}
\label{fig3}
\end{figure}

\begin{figure}[bt]
\centerline{
\epsfxsize=5cm
\epsffile{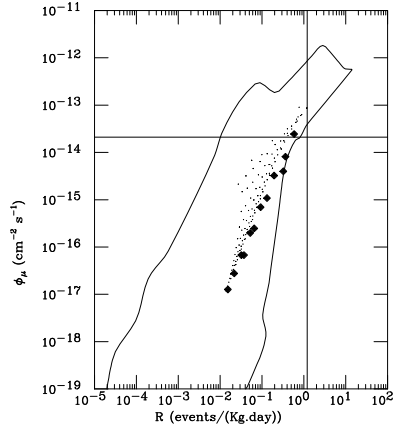}
   }
 \caption{ Comparison of the performances of currently running
 experiments of direct (Ge) and indirect (Baksan) WIMPs detection and
 neutralino configurations excluded \cite{bot40}.}
\label{fig4}
\end{figure}

From these experiments, it was concluded that the heavy Dirac neutrino is
excluded as a DM component for masses ranging from 9--10 GeV (Canfranc and
Saint Gothard), up to 4.3 TeV (Homestake and HMDM Gran Sasso). The marginal
presence of odd Ge isotopes in the Ge detectors lead to modest exclusions
for the spin dependent interactions case. Figure \ref{fig2} shows
comparatively the exclusion plots obtained for spin-independent coupling in
these experiments. Figure \ref{fig3} shows the expected
neutralino-Germanium interaction rates for some regions of the neutralino
parameter space as computed by the Torino group \cite{bot40} compared with the
Homestake background limit. The performance of these Ge detectors in direct
searches of neutralinos is compared in Figure \ref{fig4} with that
obtained in the indirect detection running experiments (for instance with
the Baksan telescope bound on the WIMP muon flux).

\subsubsection{WIMPs searches with scintillators.}

Other nuclear targets used as WIMPs scatters are sodium and iodide. The
100\% isotopic contents of A-odd isotopes of NaI scintillators ($^{127}$I
and $^{23}$Na) make them sensitive to spin-dependent interactions with WIMPs
and so significant experimental effort with sodium iodine scintillators are
underway. The level of background achieved with NaI scintillators,
has been customarily about one order of magnitude worse than that of the
best Ge detectors, as far as the bare spectra are concerned. R+D background
reduction programs of the groups of BPRS \cite{bac105} (Beijing/Paris/Rome/Saclay)
UKDMC \cite{dav106} (Imperial College/Oxford/Rutherford) and DAMA (Roma
2/Beijing/ENEA) \cite{bel125}, using selected components of very low radioactivity
have achieved levels of $\approx 1-2$ c/keV.kg.day in the relevant low
energy region (bare spectra) vs ($0.1\sim 0.2$) c/keV kg day in the best Ge
detectors. The BPRS collaboration \cite{bac105} operated in a first step (in Gran
Sasso and Frejus) NaI crystals of up to 7 kg achieving an (electron
equivalent) energy threshold of $E_{thr}=4$ keV and a background of
$B_{thr}=2$ counts/(keV kg day). Recently the DAMA \cite{bel125} group in Gran Sasso
has improved these performances with special NaI crystals of 9.7 kg. The
UKDMC group \cite{spo126}, in the Boulby mine, has worked with NaI detectors of up
to 6.2 kg, with similar remarkable results. Both collaborations have
measured the quenching factors for nuclear recoils using neutrons. Other
WIMPs searches with NaI were performed by the Zaragoza \cite{sar107} and Osaka
\cite{fus108} groups in Canfranc and Kamioka respectively by using non-dedicated
scintillators formerly employed in double beta decay searches.

\begin{figure}[bt]
\centerline{
\epsfxsize=8cm
\epsffile{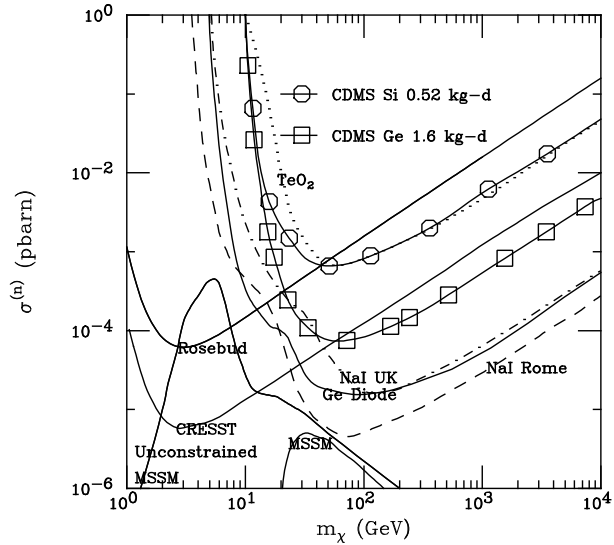}
  }
 \caption{Exclusion plots for spin-independent couplings of WIMPs
 obtained in various experiments. Expected results for sapphire bolometers
 are also depicted. Predictions of typical MSSM models are illustrated
 (see text).}
\label{fig5}
\end{figure}

\begin{figure}[bt]
\centerline{
\epsfxsize=8cm
\epsffile{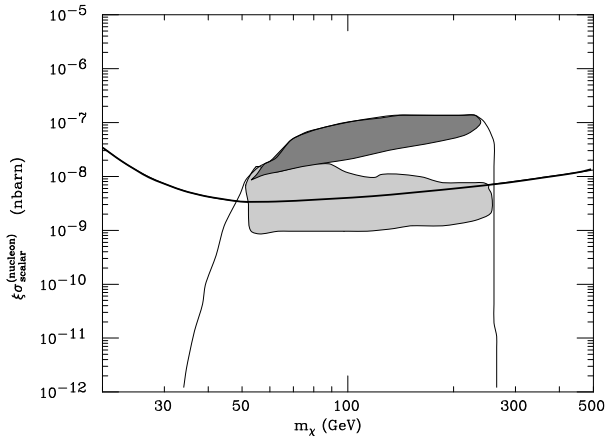}
  }
 \caption{ Regions of the neutralino parameter space which could be
 excluded \cite{bot127} by recent (NaI) spin-independent coupling
 bounds.}
 \label{fig6}
\end{figure}

The recent NaI background obtained in the UKDMC and DAMA experiments has
been approaching, and even improving that of the traditionally ultralow
background Ge diodes by incorporating pulse shape techniques of background
discrimination \cite{bel125,spo126}. To discriminate statistically the gamma
background from the nuclear recoils, the timing behaviour of the pulses
recorded for a data population---events falling in a given energy bin---is
compared with that of a template produced by external gamma and neutron
sources. It turns out that data and $\gamma $ background behavior are
essentially identical, whereas nuclear recoils have, on average, a shorter
time constant. The fraction of data which might be due to nuclear recoils
(and so to neutrons or WIMPs) is then bounded to less than 10\% to 1\%
(depending on the energy). The resulting background is accordingly reduced
from its measured level [2--4 counts/(keV kg day)] to only a few $10^{-1}$
or $10^{-2}$ counts/(keV kg day). Mass targets used are 115 kg (DAMA) and 6
kg (UKDMC) and the energy thresholds achieved in these scintillators are
2--4 keV. The new $\sigma (m)$ (cross- section WIMP-nucleon versus WIMP
mass) exclusion plots (see Figure \ref{fig5}) have already surpassed (DAMA)
those obtained from the bare spectra of Ge detectors (spin independent
interactions case) and are orders of magnitude more stringent than that of
the Ge for spin-dependent couplings. There exists some favorable regions of
the neutralino parameter space, which might be excluded \cite{bot127} by these
new spin-independent upper bounds, as depicted in Figure \ref{fig6}.

As large amounts of NaI crystals are economically affordable, sets of
scintillators have been used to explore the annual modulation of WIMPs. The
yearly modulation is originated because of the seasonal variation in the
relative velocity of the Earth and the galactic halo due to the Earth's
rotation around the Sun. The resulting net speed of the Earth with respect
to the halo reference frame oscillates between about 245 km s$^{-1}$ in
June and 215 km s$^{-1}$ in December, and so the maximum amount of energy
that can be deposited by the WIMPs in the detector, as well as their
detection rates change accordingly.

The dimensionless Earth-halo relative velocity $y(t)$ is expressed as
$y(t)=y_0+\Delta y\cos \varpi (t-t_0)$, where $y_0=1.05$ and $\Delta y=0.07$,
according to the values $v_r=232$ kms$^{-1}$, $v_{rms}=270$ kms$^{-1}$ for
the velocities, $\varpi _0=(2\pi /365)d^{-1}$, and $t_0\sim $ June 2th. The
expected signal can be expressed, at first order, as a constant term plus a
modulated component of amplitude $S_m,S(t)=S_0+S_m\cos \varpi (t-t_0)$ where
$S_m=[\partial s/\partial y]_{y_0}\Delta y$. The modulated amplitude is only
a few percent of the total DM signal. To search for the existence of a
modulation in the data, one looks for a difference between the rates of June
and December and derive the exclusion plots $\sigma (m)$ obtained by
comparing such residuals with what was to be expected. Other, more
elaborated methods of analysis which look for a possible periodic component
in the data along the whole period of data taking (like the modulation
significance method \cite{fre128}) have also been employed.

A search for annual modulation was carried out in Canfranc \cite{sar107} (Zaragoza
University) along more than two years (1993--1995) with 32 kg of NaI
scintillator corresponding to an effective exposure of 4613.6 kg day. The
data did not show any seasonal modulation but imply constraints on
cross-sections and masses of spin-dependent and spin-independent
interactions of WIMPs more stringent than that obtained from the customary
method of comparing the total expected rate with that experimentally
measured. A larger experiment with seven NaI scintillators (a total of 75
kg), with upgraded PMs which use light guides and low background components
is being installed at Canfranc (2450 m.w.e.), in a shielding with
archaeological lead. Early experiments on WIMP signal modulation were also
carried out with Ge detectors \cite{sar124}.

A liquid Xenon scintillator \cite{bel110} is being used by the DAMA Collaboration at
Gran Sasso. An experiment with 6.5 kg. of liquid Xe ($\sim $2 liters)
isotopically enriched to 99.5\% in $^{129}$Xe with a background of $B\approx
3$ counts/(keV kg day) at threshold $E_{thr}\approx 10$ keV is looking for
annual modulation. Preliminary results corresponding to an exposure of 31.4
days in winter and 66.1 days in summer did not find seasonal effects.

After these pioneering searches for WIMP annual modulation signals with NaI
and liquid Xe scintillators, another experiment with a large mass of NaI
scintillators (nine NaI crystals of 9.7 kg each) is presently running at
Gran Sasso operated by the DAMA collaboration in a quest for seasonal
modulations. The first reposted results \cite{bel129}, corresponding to an
statistics of about 39 days in winter and 14 days in summer, bring forward
an indication of a possible modulation effect $S_m=0.034\pm 0.008$
count/(keV kg day) in the 2--12 keV region, as analyzed by the modulation
significance method. The analysis is extended to the whole relevant energy
interval by using a maximum likelihood method to conclude that the reported
annual modulation might be interpreted as due to a 60 GeV WIMP with an
interaction cross-section on protons of $10^{-5}$ pb. The (preliminary)
results of this work, however, do not allow any final statement on the
presence of a signal and a larger statistics is needed, together with
further effort in stability monitoring. Critical comments on these results
can be found in \cite{ger130}. The data taking of the DAMA experiment
continues with the purpose of congiguring the modulation hint shown in
the preliminary data. Possible implications of
these data for relic neutralinos have been analyzed by the
Torino group \cite{bot131}.

\subsubsection{Searches with other techniques}

There exist much more activity and projects on WIMP direct detection with
conventional detectors. For instance, calcium fluorine scintillators are
being explored for dark matter experiments by the groups of Roma 2, Osaka
and Milan. The detector ELEGANT VI (Osaka) \cite{haz132} consisting of CaF$_2$(Eu)
scintillators (a total of 3.5 kg of $^{19}$F), surrounded by CsI detectors
as an active shield, is looking for WIMPs scattering on $^{19}$F (which has
a favorable nuclear factor of merit for DM searches). After a preliminary
test run at sea level, the set of calcium fluorine rods will go underground
in the Otho Laboratory. On the other hand, the UCLA-UKDM collaboration is
investigating the WIMP detection with liquid Xenon detectors \cite{sum133}. Nuclear
recoil identification can be achieved in liquid Xenon either by
scintillation pulse shape analysis or by drifting the ionization to produce
a second proportional scintillation pulse. The ratio between the primary
(liquid Xenon direct scintillation) and secondary pulses is different for
electron and nuclear recoils and so the background can be efficiently
discriminated \cite{ben134}. A 2-kg ICARUS-WIMP liquid Xenon device is operating at
the Montblanc underground laboratory. A modular prototype of such
ICARUS-like detector, called ZEPLIN, with 20 kg of liquid Xe, will be
installed at Boulby \cite{par135}.

Prototype detectors of moderately superheated microdroplets \cite{zac136,col137}
are being installed. Superheated Droplets Detectors (SDD) consist of a
dispersion of droplets ($\emptyset \sim ~10-100$ $\mu $m) (say of Freon-12)
of a superheated liquid fixed in a viscous polymer or aqueous gel. These
devices are insensitive to energetic muons, gamma rays, X-ray and beta
particles, while responding to neutron recoils of only a few keV. The energy
deposition in the droplet produces a phase transition from superheated to
normal, causing its vaporization in a bubble of $\emptyset \sim 1$ mm, which
can be optically recorded or detected with piezo-electric sensors. An
experiment (SIMPLE) is being installed by a Paris Univ. VII/CFN Lisbon Univ.\
collaboration \cite{col138} at a shallow depth near Paris, whereas a prototype
projected by the Montreal/Chalk River collaboration \cite{ham139} is underway.

Large TPCs \cite{ger140} for DM are also in the prospective program of various
groups (Saclay and San Diego), whereas the MUNU experiment \cite{bro141} will
provide a device for developing methods in DM searches. In fact, low
pressure gas detectors have been receiving attention for a long time because
of their potential sensitivity to the nuclear recoil direction \cite{ric142},
a distinctive signature for WIMP identification \cite{spe113}. Direction-sensitive
mechanisms and detectors have been proposed, among them, we mention the use
of organic crystals (anthracene) \cite{bel143}, the ejection of atoms from surfaces
\cite{col144}, the detection of the recoiling silicon atoms in the surface layers of
silicon chips by means of thin film thermal detectors \cite{mar145} and the rotons
in superfluid helium \cite{ada146}. Imprints left by WIMPs in old ancient mica are
also being searched for \cite{sno147}, taking advantage of the fact that over long
periods of time, WIMPs should have collided with underground mica ($^{16}$O,
$^{28}$Sr, $^{27}$Al, $^{39}$K) and left tracks.

The Heidelberg/Moscow Collaboration is planning the use of a pure Ge
detector, inside a set of Ge detectors. Preliminary results on background
reduction obtained with a Ge detector embedded in a well of Ge in the same
cryostat have already been obtained \cite{bau148}. This collaboration is planning a
large dark matter and double beta decay experiment (Germanium Nitrogen
Underground Search, GENIUS) with a cooled array of 300 enriched Ge detectors
of a mass of 1 ton immersed in liquid nitrogen. We refer the reader
to the recent reviews and topical Workshops on the subject \cite{tau150,she151}.

\subsubsection{Cryogenic Detectors \cite{cry115,sup116,fio117,boo114,sad152}.}

In the WIMP scattering on matter, only a small fraction of the energy
delivered by the WIMP goes to ionization, the main part being released as
heat. Consequently, thermal detectors (quenching factor close to one) should
be suitable devices for dark matter and other rare event searches. Moreover,
the mechanisms and quanta involved in the detection imply that such
detectors should have better energy threshold and energy resolution than
conventional detectors. As a bonus, they will allow to enlarge the number of
target nuclei (with various spin and nuclear masses) for WIMPs interactions.
An important property is that they permit to discriminate backgrounds by
simultaneously measuring the heat and the ionization (or scintillation).
Some examples of cryogenic detectors are: bolometers (with various types of
sensors); superconducting tunnel junctions; superheated superconducting
grains (SSG), quasiparticle detectors, and cryogenic hybrid devices (i.e.
detectors which are sensitive to both the heat and the ionization produced
by the DM particle). Two types of cryogenic
detectors are being developed: bolometers \cite{fio153} (either pure thermal or
thermal plus ionization hybrid detectors) and superconducting superheated
grains \cite{pre154,sup116,way155}. The development of bolometers along the
past few years has been impressive.

Bolometers measure the increase of temperature produced by the recoiling
nucleus hit by a WIMP, which is proportional to the ratio between the energy
thermally released and the heat capacity of the crystal $\Delta T=\Delta
E/CV $, where $C$ is the heat capacity, and $V$ the volume of the bolometer.
The temperature pulse is detected with a sensor in thermal contact with the
absorber. An appealing feature of these detectors is that their energy
resolution should be a priori much better than that of conventional
detectors, as the energy deposition mechanism is made in terms of phonons
($10^{-4}-10^{-6}$ eV). The energy resolution achieved is damped by several
effects, but in the keV region, resolutions of the order of 100 to 300 eV
have been achieved even for large mass crystals. Dielectric and diamagnetic
materials---for which the heat capacity is proportional to the cube of the
working to the Debye temperature ratio---have been widely used. That makes
necessary to employ materials with large Debye temperature and work at
temperatures as low as possible (of the order of tens of mK). The
$^3$He/$^4$He dilution refrigerators (DR) needed for such searches, are not optimized
for low background and so much work is being devoted recently to provide in
the DR components the low radioactivity environment needed in rare event
searches. Sensors used are neutron transmutation doped (NTD) thermistors
(glued or bounded to the crystal), superconducting phase transition
thermometers (SPT) and quasiparticle trapping assisted electrothermal
feedback transition edge thermometers (QET).

The first bolometer operating underground (Gran Sasso) was that of the Milan
group dedicated to double beta decay searches. Large bolometers of TeO$_2$
(334 g) with NTD Ge-sensors (Milan group, Gran Sasso), (resolution of 1\% at
60 keV and background of $\approx 12$ counts/(keV.kg.day) at threshold,
$E_{thr}\approx 10$ keV), although optimized for $2\beta $ decay searches
\cite{ale156}, have produced data for constraining dark matter particle
\cite{ale157} (see Figure \ref{fig5}) and have proved the sensitivity of
these bolometers to nuclear recoils \cite{ale158}. The nuclear recoil quenching factor
Q of these bolometers has been measured by using two 73 g tellurium oxide
crystals at 22 mK, mounted face to face and measuring the energy of the $\alpha $
and recoils signals produced by an implanted $\alpha $ source \cite{ale158}. The
electron response was calibrated with $\gamma $ sources. The result is
$Q_{recoil}=1.025\pm 0.01(stat)\pm 0.02(syst)$, practically independent of
energy (from 10 to 200 keV). The current set-up of this experiment, four
crystals of 334 g each, has been recently extended successfully to a set of
20 crystals. A new project to be installed in Gran Sasso, named CUORE
(Cryogenic Underground Observatory for Rare Events) is a large extension of
the Milan cryogenic array of twenty 340 g TeO$_2$ crystals with NTD Ge glued
thermistors. CUORE\ \cite{fio159} will consists of an array of one thousand crystals
of TeO$_2$ (750 g each) with NTD Ge sensors, cooled down to 10 mK, planned
for double beta decay, direct detection of WIMPs, solar axion searches, low
energy nuclear physics and, possibly, interactions of antineutrinos from
artificial sources. Other absorbers can be also considered, like Al$_2$O$_3$,
PbWO$_4$, Ge, CaF$_2$, but the first step of CUORE---called
CUORICINO---will choose the TeO$_2$ option. CUORICINO will consist of 100
crystals of TeO$_2$ of 5 cm side with a total mass of 75 kg, i.e. about 20
kg of $^{130}$Te, an amount by far larger than in any running double beta
decay experiment. That will allow a very significant experiment on double
beta decay of $^{130}$Te, as well as on WIMP annual modulation search.

The EDELWEISS (Experiment pour Detecter les WIMPs en Site Souterraine)
Collaboration (Saclay/ IAS/Orsay/College de France/IAP/Lyon/Modane) has
pioneered the operation at Frejus of a bolometer experiment dedicated to DM
searches using a 24 g sapphire crystal endowed with an NTD Ge thermistor
\cite{cor160}. Recent improvements of this experiment, have produced energy
resolutions of $\approx 3-4$ keV at 60 keV, thresholds of 1--2 keV and
backgrounds of 25 counts/(keV.kg.day) at low energy \cite{yvo161}.

The CRESST (Cryogenic Rare Event Search with Superconducting Transition
Thermometers) Collaboration (Munich/Garching/Oxford) has developed sapphire
bolometers with superconducting phase transition thermometers SPT (in
indium, indium/gold and tungsten). This Collaboration got \cite{sei162}, with
a 31 g sapphire bolometer (at 15 mK), an energy threshold of 0.3 keV and an energy
resolution of $\approx 100$ eV for 1.5 keV X-rays FWHM, which is the best
resolution obtained so far per unit of detector mass. The CRESST experiment
\cite{sis163} employs four sapphire bolometers of 262 g each with tungsten SPT . The
energy resolution is 250 eV at 1.5 keV and the expected energy threshold is
500 eV. After a successful running test at Gran Sasso, a new improved set-up
(with selected ultra-low background materials, cold box, copper frame) is
being implemented. The CRESST predicted exclusion plot for a flat background
of 1 count/(keV kg day) is also shown in Figure \ref{fig5}. The MSSM
predictions (see for instance Refs. \cite{jun6} and \cite{gab39}) are also
depicted to illustrate the sensitivity of these searches for neutralino
dark matter.

On the other hand, small sapphire bolometers of $E_{thr}\approx 0.3$ keV and
energy resolution of 120 eV at 1.5 keV made by IAS (Orsay), are being
explored in Canfranc at 2450 m.w.e. The ROSEBUD (Rare Objects Search with
Bolometers Underground) \cite{bob164} experiment (IAS/IAP/Zaragoza) consists
of two sapphire crystals of 25 g (and 50 g) each, cooled at 20 mK in a small
DR shielded inside and outside with archaeological lead. Anticipated upper
bounds for a background of 1 count/(keV kg day) are given in Figure
\ref{fig5}.

A further step in the background discrimination has been recently achieved
(CDMS \cite{nam165} and EDELWEISS \cite{hot166}) by measuring simultaneously
the heat and ionization produced in the detector by the WIMP-nucleus scattering.
The energy released by the recoiling nucleus impinged by the WIMP (or the
recoiling electron impinged by a particle) appears in form of phonons and
electron-hole pairs. Simultaneous measurement of both quantities for each
event allows the discrimination of electron recoil events (tracers of the
background) from nuclear recoil events (WIMP signal plus neutrons) because
for a given deposited energy---measured as phonons---the ionization produced
by the recoiling nucleus is less than that generated by electrons.

Low temperature hybrid devices \cite{fio117,sad152} have been developed by the CDMS
(Cryogenic Dark Matter Search) \cite{shu167} collaboration (UC Berkeley / CfPA /
UCSB/ Stanford) with Ge bolometers which also collect electron-holes pairs.
The proof-of-principle of a 70 g. hybrid Ge detector was successful and good
discrimination efficiency between nuclear recoils and Compton background was
obtained. Two BLIP \cite{gai168} (Berkeley Large Ionization and Phonon based) Ge
detectors (62 g and 165 g) with NTD Ge bounded thermistors have been
operated at shallow depth in Stanford in the low background cold box of a DR
at 20 mK. In another type of detector, called FLIP \cite{dav169} (Fast Large
Ionization and Phonon-based), the phonon sensors used are QETs
(Quasiparticle trapping assisted Electrothermal feedback Transition edge
thermometers) with aluminium phonon collector pads which absorbe the phonons
by breaking copper pairs and forming quasiparticles, which are trapped in
tungsten. This trapping produces an increase in resistance showed up as a
current pulse with a SQUID array. Both Ge-BLIP and Si-FLIP detectors have
already produced results.

The BLIP 62 g Ge detector, for instance, has a gamma rejection ratio higher
than 99\% and a nuclear acceptance ratio greater than 95\%. The FLIP 100 g
Si detector has, respectively, the ratios 99\% and 75\% (at low
energies). The phonon and ionization energy resolution FWHM of the 62 g Ge
BLIP are respectively 500 eV and 1.5 keV. The recoil energy threshold is
about 15 keV. The nuclear recoil background is already as low as 0.3
count/(keV kg day). The 165 g Ge BLIP has energy resolutions of about 1 keV
FWHM in both channels, whereas the threshold is $\sim 2$ keV. On the other
hand, the FLIP detector has 7 keV FWHM as phonon energy resolution and 2 keV
as ionization energy resolution. The recoil energy threshold in this
detector is $\sim 30$ keV. The background rate recorded in the above BLIP
and FLIP detectors in anti-coincidence with the muon-veto stands,
respectively, at about 4 counts/(keV kg day), 2 counts/(keV kg day) and 6
counts/(keV kg day). The corresponding nuclear recoil background---obtained
after performing the phonon-ionization discrimination of events quoted
above---goes down to 0.1--0.3 count/(keV kg day). These remarkable
results give confidence in the method and techniques followed in the quest
of the neutralino sensitive region. A tritium contamination is being
removed. Problems concerning the drop in the discrimination efficiency at
low energies due to the dead layer in the Ge detectors are being addressed.
The exclusion plots obtained up to now by this Collaboration are shown in
Figure \ref{fig5}, compared with those obtained in other running
experiments. A next step will be the operation of these detectors in the
SOUDAN underground facility.

The EDELWEISS Collaboration (Frejus) is now developing a 70 g Ge bolometer
with simultaneous detection of heat and ionization, obtaining similar
results to those of the BLIP Ge detector. An energy threshold of 4 keV (both
phonons and ionization), an energy resolution of 1.25 keV FWHM for the
phonon channel and of 1.3 keV for the ionization channel, both at $E=122$
keV have been achieved \cite{hot166}. A good separation of the neutron and gamma
events has been obtained. A rejection efficiency of 98\% for neutron events
from a source of $^{232}$Cf has been achieved relative to the gamma
background of a $^{57}$Co source. This discrimination efficiency has allowed
to reduce the effective background which could be attributed to WIMPs
(nuclear recoils) to levels with an upper limit of 0.5 count/(keV kg day) in
the 15 to 45 keV nuclear recoil energy. The preliminary results at a short
running are reported in Ref. \cite{hot166}.

The proof-of-principle of simultaneous measurements of heat and light was
made by the Milan group with a small (2g) CaF$_2$ scintillator bolometer and
by the Lion and Tokyo groups with small LiF crystals, but these
scintillating bolometers are still in a R+D stage.

The Superheated Superconducting Granules (SSG) detector proposal is about
thirty years old \cite{ber172}. Proposed originally as neutrino detectors, they were
then extended to other rare event processes like double beta decay or
particle DM\ searches \cite{ale156,pre154,way155}. The signal produced in the
SSG detectors,
as response to the particle interaction, is due to the disappearance of the
Meissner effect when the heat delivered by the particle energy deposition is
able to trigger the superconducting to normal state phase transition (flip)
of the grains. A suitable array of coils, embedded in the colloid measures
the voltage drop produced by the local change of the magnetic flux over the
flipped grains or region, in the external applied magnetic field. The SSG
offer a fast (ns) timing capability; a unique background rejection (97\%)
(since only a single grain is expected to flip per WIMP interaction, in
contrast to several grains with standard ambient radiation fields); a
sensitivity to very low energy deposition (as proved in neutron irradiation
experiments); and the advantage of having the readout and the SSG device
thermally decoupled. The main disadvantage is the small instrumented mass
and, consequently, the need for large-scale electronics. That small mass
constraint is due to the small filling factor currently employed in these
detectors, i.e. the majority of a SSG device is inert.

In particular, the Bern/PSI/Annecy group is constructing a Superconducting
Superheated Grains (SSG) prototype detector (ORPHEUS experiment) \cite{abp170},
with 100 g of Sn (to be extended to 1-kg, and other targets) micrograins of 5 to
20 $\mu $m of diameter, to be operated at shallow depth. This group has
proved the sensitivity of SSG detectors to nuclear recoils of only a few keV
produced by elastic scattering of the PSI 70 MeV neutron beam with
micrograins of Al (23 $\mu $n), Zn (19 $\mu $n) and Sn (17 $\mu $). Phase
transitions were also observed when these devices were exposed to
radioactive sources or to the plain background. The Lisbon-Zaragoza-Paris
Collaboration \cite{gir171} is installing in Canfranc a pilot experiment (SALOPARD)
also with an SSG suspension of 100 g of tin micrograins of 20 $\mu $m
diameter. The gamma rejection predicted is about 95\%, and the expected
energy threshold of $\sim 2$ keV. This group has proved that these devices
can produce energy spectra by suitable sweeping of the applied magnetic
field. Improvements in the radiopurity of the SSG detectors is a main
question still to be solved by ORPHEUS and SALOPARD.

Other type of Superconducting detectors, like the Superconducting Tunnel
Junction (STJ), made very important R+D progresses following the
developments of the quasiparticle trapping technique \cite{boo173} of the Oxford
group, but there is not yet any planned STJ dark matter experiment.

\section{Prospects of Future Experiments}

The quest for the particle dark matter faces formidable tasks. To fulfill
the requirements implied by the strategies indicated above, improvements in
the radioactive background (both intrinsic and environmental), in detector
efficiency and in energy threshold should be accomplished. The use of
radiopure material need to be implemented by that of background discrimination
techniques. New, suitable nuclear targets should be also tried. The search
for modulated, distinctive signatures of WIMPs should be pursued. Most of
these requirements will be hopefully fulfilled by sets of cryogenic/hybrid
detectors, superconducting detectors or large masses of NaI scintillators or
of other conventional detectors endowed with background discrimination, and
this line of action is in the objective of the forthcoming or future
experiments.

\subsection*{Acknowledgments}
I am indebted to Eduardo Garc\'{\i}a and Stefano Scopel for their cooperation
in the making of the exclusion plots, and CICYT (Spain) for
financial support.


\begin{thebibliography}{999}
\bibitem{fab1} S.M. Faber andJ.S.Gallager, Ann. Rev. Astron. Astrophys.
17 (1979) 135.
\bibitem{pri2} J.R. Primack, D. Seckel and B. Sadoulet,
Ann. Rev. Nucl. Part. Sci. 38 (1988) 751.
\bibitem{tri3} V. Trimble, Ann. Rev. Astron. Astrophys 425 (1989)
25.
\bibitem{kol4} E. Kolb and M. Turner, {\em The Early Universe}, Addison Wesley, N.Y.,
(1990).
\bibitem{fic5} M. Fich and S. Tremaine, Ann. Rev. Astron. Astrophys
409 (1991) 29
\bibitem{jun6} G. Jungman, M. Kamionkowski and K. Griest, Phys. Rep. 267
(1996) 195
\bibitem{pri7} J.R. Primack, in: Proc. Int. School on Dark Matter in the Universe,
(Varenna) (IOS Press) ed. S. Bonometo et al. Preprint astro-ph/9604184 \\
J. Primack, Proc. Int. Conference on Neutrino Physics and Astrophysics, Neutrino 96
(Helsinki, June 1996), ed. K. Enqvist et all, (World Scientific, 1997) p. 398
and references therein.
\bibitem{str8} M. Strauss and J. Willick, Phys.Rep. 261 (1995) 271.
\bibitem{bah9} N. Bahcall et al. Ap. J. 447 (1995) L81.
\bibitem{fre10} W.L. Freeman et al., Nature 371 (1994) 757.
\bibitem{bir11} M. Birkinshaw, J.P. Hughes, Ap.J. 420 (1994) 33.
\bibitem{pee12} P.J.E. Peebles., {\em Large Scale Structure of the Universe}, Princeton University
Press, 1980.
\bibitem{vit13} N. Vittorio, R. Juszkiewicz, {\em Large Scale Structure and Motions in the Universe}, ed.
M. Mezzoti et al. (Kluwer) p. 241.
\bibitem{row14} M. Rowan-Robinson, MNRAS, 1993 and Pre. Nat. Ac. Sci. 90 (1993) 4822.
\bibitem{ber15} E. Bertschinger, A. Dekel, Ap.J. Lett. 336 (1989) L5.
\bibitem{dek16} A. Dekel et al., preprint IASSNS-AST 92/55. \\
A. Dekel, Rev. Astron. Astrophys. 32 (1994) 371.
\bibitem{per17} S. Perlmutter et al., preprint astro-ph/9602122; LBL-39291.
\bibitem{smo18} G.F. Smoot et al., Astrophys. J. 396 (1992) L1.
\bibitem{wal19} T.P. Walker et al. Ap.J. 376 (1991) 51. \\
C.J. Copi, D.N.Schramm and M.S. Turner, Phys. Rev. Lett. 75 (1995) 3981.
\bibitem{sha20} Q. Shafi and F. W. Stecker, Phys. Rev. Lett. 53 (1984) 1292.
\bibitem{blu21} G.R. Blumental et al., Nature 311 (1994) 517. \\
J. Primack and A. Klypim, in Proc. Conference on Sources and Detection of Dark
Matter in the Universe, (UCLA, Feb. 1996), Nucl. Phys. B (Proc. Suppl.) 1996.
\bibitem{pog22} D. Pogosyan and A. Starobinsky, Ap. J. 447 (1995) 465. \\
J.R. Primack et al. Phys. Rev. Lett. 74 (1995) 2160.
\bibitem{tur23} M. Turner, ``Inflation after COBE'', Fermilab Conf 92/313A and
``Dark Matter Theoretical Perspectives'', Fermilab Conf. 92/382A. \\
E.L. Wright et al., Ap. J. 396 (1992) L13. \\
R.K. Schaefer and Q. Shafi, Nature 359 (1992)199.
\bibitem{pac24} B. Paczynski, Ap.J. 304 (1986) 1.
\bibitem{alc25} C. Alcock et al., Nature 365 (1993) 621; Phys. Rev. Lett. 74 (1995) 2867;
Ap.J. 461(1996)89; preprint astro-ph/9606165.
\bibitem{aub26} E. Auburg et al., Nature 365 (1993) 623. \\
C. Renault et al., Astron. Astrophys. 324 (1997) L69.
\bibitem{uda27} A. Udalski et al., Acta Astronomica 44 (1994) 165.
\bibitem{ber28} See for instance V. Berezinsky et al. Nucl. Phys. B.(Proc. Suppl.)
48 (1996) 22. \\
J. Ellis, Nucl. Phys. B (Proc. Suppl.) 48 (1996) 522.
\bibitem{dol29} See for instance A.D. Dolgov, Nucl. Phys. B (Proc. Suppl.) 48 (1996) 5 and
references therein.
\bibitem{dum30} J. Dumarchez et al., (NOMAD Coll.), in Proc. Conference on Neutrino Physics
and Astrophysics, Neutrino 96, (Helsinki, June 1996), ed. by K. Enqvist et al.,
(World Scientific, 1997) p. 143. \\
D. Macina et al., (CHORUS Coll.), Nucl. Phys. B (Proc. Suppl.) 48 (1996) 183.
\bibitem{lee31} B.W. Lee and S. Weinberg, Phys. Rev. Lett. 39 (1977) 165.
\bibitem{ell32} J. Ellis et al., Nucl. Phys. B 238 (1984) 453.
\bibitem{hab33} H.E. Haber and G.L. Kane, Phys. Rep. 117 (1985) 75.
\bibitem{gri34} K. Griest, M. Kamionkowski and M.S. Turner, Phys. Rev. D 41 (1990)
3565.
\bibitem{sre35} M. Srednicki and R. Watkins, Phys. Lett. B 225 (1989) 140.
\bibitem{giu36} G.F. Giudie and E. Roulet, Nucl. Phys. B 316 (1989) 429.
\bibitem{dre37} M. Drees and M.M. Nojiri, Phys. Rev. D48 (1993) 3483.
\bibitem{ell38} J. Ellis, Nucl. Phys. B (Proc. Suppl.) 35 (1994) 5. \\
L. Roszkowsky, Phys. Lett. B 278 (1992) 147.
\bibitem{gab39} A. Gabutti et al., Astropart. Phys. 6 (1996) 1.
\bibitem{bot40} A. Bottino et al., Mod. Phys. Lett. A7 (1992) 733; Phys. Lett.
B 380 (1996) 308. \\
S. Scopel, in Proc. Workshop on Dark Matter in Astro and Particle Physics
(Heidelberg, September 1996), eds. H.V. Klapdor and Y Ramachers  (World Scientific,
1997) p. 358.
\bibitem{bot41} A. Bottino et al., Astropart. Phys. 2 (1994) 77.
\bibitem{gri42} K. Griest, Phys. Rev. D 38 (1988) 2357; Phys. Rev. Lett.
61 (1988) 666.
\bibitem{pec43} R. Peccei and H.R. Quinn, Phys. Rev. Lett. 38 (1977) 1440.
\bibitem{tur44} M. Turner, Phys. Rep. 197 (1990) 67.
\bibitem{raf45} G. Raffelt, Phys. Rep. 198 (1990) 1; in Proc. Moriond Workshop,
Dark Matter in Cosmology (January 1995), preprint hep-ph/9502358.
\bibitem{sik46} P. Sikivie, Phys .Rev. Lett. 51 (1983) 1415; Int. J. Mod. Phys.
D 35 (1994) 1.
\bibitem{hag47} C. Hagman et al., Phys. Rev. D 42 (1990) 1297.
\bibitem{wue48} W. Wuenshch et al., Phys. Rev. D 40 (1989) 3153.
\bibitem{bib49} K. van Bibber et al., Phys. Rev. D 39 (1989) 2089; preprint UCRL-JC-118357;
and Int. J. Mod. Phys. D 35 (1994) 33.
\bibitem{oga50} I Ogawa, S. Matsuki and K. Yamanoto, Phys. Rev. D 53 (1996) 1740.
\bibitem{sem51} Y.K. Semertzidis et al., Nucl. Instrum. Methods Phys. Res. A 356
(1995) 122.
\bibitem{hag52} C. Hagmann et al., Lawrence Livermore Nat. Lab. Preprint, 1996. \\
E.J. Daw and K. van Bibber, in Proc. Workshop on the Identification of Dark Matter,
(Sheffield, September 1996), ed. N. Spooner (World Scientific, Singapore.
1997) p. 362.
\bibitem{che53} S.L. Cheng, C.Q Geng and W.T. Ni, preprint NUCU-HEP-94-20 and
hep/ph/9508013.
\bibitem{raf54} G. Raffelt, A. Werss, Phys. Rev. D 51 (1995) 1495.
\bibitem{kra55} S.V. Krasmikov, Phys. Rev. Lett. 76 (1996) 2633.
\bibitem{jan56} H.T. Janka et al., Phys. Rev. Lett. 76 (1996) 2621.
\bibitem{laz57} D.M. Lazarus et al., Phys. Rev. Lett. 69 (1992) 2333.
\bibitem{buc58} W. Buckmiller and F. Hoogeveen, Phys. Lett. B 237 (1990) 278.
\bibitem{pas59} E.A. Paschos and K. Zioutas, Phys. Lett. 323 (1994) 367.
\bibitem{cre60} R.J. Creswick et al., submitted to Phys. Lett. B (1997).
\bibitem{avi61} F.T. Avignone et al., submitted to Phys. Lett. B (1997).
\bibitem{smi62} P.F. Smith and J.D. Lewin, Phys. Rep. 187 (1990) 203.
\bibitem{sil63} J. Silk and M. Srechniki, Phys. Rev. Lett. 53 (1984) 624; \\
M. Kamionkowski, Phys. Rev. D 44 (1991) 3021; \\
M. Kamionkowski, in Particle Astrophysics, Proc. Moriond Workshop (January 1994),
ed. J. Tran Tanh Van (Editions Frontieres, 1994). \\
A. Bottino et al., Astrop. Phys. 3 (1995) 77.
\bibitem{sal64} M. Salamon et al., Ap. J. 349 (1990) 78. \\
R. E. Streitmatter et al., Proc. 21st Int. Cosmi Rays Conference, (Adelaide, 1990)
OG7.3-2, p. 277.
\bibitem{ahl65} S. Ahlen et al., Nucl. Instrum. Methods Phys. Res. A 350 (1994) 351.
\bibitem{adr66} O. Adriani et al. in Proc. 24th Int. Cosmic Rays Conference (Roma 1995)
 OG v3, p. 591.
\bibitem{gou67} A. Gould, Ap.J. 321 (1987) 560; Ap.J. 388 (1991) 338.
\bibitem{sil68} J. Silk, K. Olive and M. Srednicki, Phys. Rev. Lett.
55 (1988) 257.
\bibitem{kra69} L.M. Krauss, K. Freese, D.N. Spergel and W.H. Press,
Astrophys. J. 299 (1985) 1001.
\bibitem{kra70} L.M. Krauss, M. Srednicki and F. Wilczek, Phys. Rev. D 33 (1986) 2079.
\bibitem{fre71} K. Freese, Phys. Lett. B 167 (1986) 295.
\bibitem{gai72} T. Gaisser, G. Steigman, and S. Tilav, Phys. Rev.
D 34 (1986) 2206.
\bibitem{bot73} A. Bottino et al., Astropart. Phys. 3 (1995) 65.
\bibitem{hal74} F. Halzen et al., Phys. Rev. D 45 (1992) 4439.
\bibitem{ber75} V. Berezinsky et al., preprint CERN-TH-96-42.
\bibitem{ber76} L. Bergstrom, J. Edsj\"{o} and P. Gondolo, preprint hep-ph/9607237; \\
L. Bergstrom, in Proc. Conf. on Neutrino Physics and Astrophysics, Neutrino 96,
(Helsinki, June 1996), eds. K. Enqvist et al. (World Scientific, 1997) p. 436.
\bibitem{hal77} F. Halzen, Nucl. Phys. B (Proc. Suppl.) 38 (1995) 472. \\
T.K. Gaisser, F. Halzen and T. Stanev, Phys. Rep. 258 (1995) 173. \\
F. Halzen, in Proc. Workshop on Topics in Astroparticle and Underground
Physics, TAUP 97 (Gran Sasso, September 1997), eds. A. Bottino et al., to appear in
in Nucl. Phys. B (Proc. Suppl.) 1998.
\bibitem{mon78} T. Montarulli et. al., (MACRO Coll.) Nucl. Phys. B (Proc. Suppl.)
48 (1996) 87.
\bibitem{bol79} M. Boliev, Nucl. Phys. B (Proc. Suppl.) 48 (1996) 83.
\bibitem{los80} J. M. Losecco et al., Phys. Lett. B 188 (1987) 388. \\
R. Svoboda et al., Astrophys. J. 315 (1987) 420.
\bibitem{mor81} M. Mori et al., Phys. Lett. B 205 (1988) 406; Phys. Lett.
B 278 (1991) 217; Phys. Lett. B 289 (1992) 463; Phys. Rev. D48 (1993) 5505.
\bibitem{bez82} L. B. Bezrukov et al., in Proc. Workshop on The Dark Side of the
Universe (Roma, November 1995), ed. R. Bernabei (World Scientific, Singapore
1996). \\
L.A. Belolaptikov et al., in Proc. Conference on Neutrino Physics and Astrophysics,
Neutrino 96 (Helsinki, June 1996), eds. K. Enqvist et al. (World Scientific,
Singapore 1997) p. 524.
\bibitem{bar83} S.W. Barwick, Nucl. Phys. B (Proc. Suppl.) 43 (1995).
\bibitem{res84} L. Resvanis, Nucl. Phys. B (Proc. Suppl.) 48 (1996) 425.
\bibitem{lea85} J.G. Learned, Nucl. Phys. B (Proc. Suppl.) 31 (1993) 484;
Nucl. Phys. B (Proc. Suppl.) 38 (1995) 480.
\bibitem{mos86} L. Moscoso, ``Neutrino Astronomy'', in Proc. Rencontres de Moriond
(January 1997), preprint DAPNIA/SPP 97-05.
\bibitem{nut87} OECD Megascience Forum Workshop on A 1Km3 Deep Sea
Neutrino Observatory (Taormina, Sicily, June 1997); \\
Series of ``Neutrino Telescopes'' Workshops,
Venezia, ed. M. Baldo Ceolin, Pub. Istituto Venetto 1990-1996,
\bibitem{bos88} P.C. Bosetti, Nucl. Phys. B (Proc. Suppl.) 48 (1996) 466.
\bibitem{hul89} P.O. Hulth et al., (AMANDA Coll.), in Proc. Conference on Neutrino
Physics and Astrophysics (Helsinki, June 1996), eds. K. Enqvist et al. (World
Scientific, Singapore, 1997) p. 518.
\bibitem{mon90} F. Montanet (ANTARES Coll.), in Proc. Winter Meeting on
Fundamental Physics (Formigal,Spain, March 1997), eds. M. Aguilar et al.
(World Scientific, Singapore, 1998; \\
ANTARES Proposal CCPM-97-02, DAPNIA 97-03.
\bibitem{spi91} Ch. Spiering et al., (Baykal Coll.) Nucl. Phys. B (Proc. Suppl.)
48 (1996) 463.
\bibitem{mon92} B. Monteleoni et al., (NESTOR Coll.), in Proc. Conference on
Neutrino Physics and Astrophysics (Helsinki, June 1996), eds. K. Enqvist et al.
(World Scientific, Singapore, 1997) p. 534.
\bibitem{goo93} M.W. Goodman and E. Witten, Phys. Rev. D31 (1986) 3059.
\bibitem{was94} I. Wasserman, Phys. Rev. D33 (1986) 2071.
\bibitem{ber95} R. Bernabei, Riv. Nuovo Cimento 18 (1995) 1.
\bibitem{avi96} F.T. Avignone and A. Morales, in Proc. Conference on Neutrino
Physics and Astrophysics, Neutrino 96, (Helsinki, June 1996),  eds. K. Enqvist
et al. (World Scientific, Singapore, 1997) p. 413.
\bibitem{cal97} D.O. Caldwell, in ``Direct searches for relic particles'',
Proc. Workshop on Topics in Astroparticle and Underground Physics, TAUP 97,
eds A. Bottino et al., to appear in Nucl. Phys. B (Proc. Suppl.) (1998).
\bibitem{ahl98} S.P. Ahlen et al., Phys. Lett. B195 (1987) 603.
\bibitem{dru99} A.K. Drukier et al., Nucl. Phys. B (Proc. Suppl.)
28A (1992) 293.
\bibitem{cal100} D.O. Caldwell et al., Phys. Rev. Lett. 61 (1988) 510.
\bibitem{reu101} D. Reusser et al., Phys. Lett. B 255 (1991) 143.
\bibitem{mor102} J. Morales et al., Nucl. Instrum. Methods Phys. Res. A 321 (1992) 410; \\
E. Garc\'{\i}a et al., Phys. Rev. D 51 (1995) 1460.
\bibitem{bec103} M. Beck et al., Phys. Lett. B 336 (1994) 141.
\bibitem{cal104} D. O.Caldwell, Nucl. Phys. B (Proc. Suppl.) 38 (1995) 394.
\bibitem{bac105} C. Bacci et al., Phys. Lett. B 293 (1992) 460. \\
A. Bottino et al., Phys. Lett. B 295 (1992) 330.
\bibitem{dav106} G.J. Davies et al., Phys. Lett. B 322 (1994) 159. \\
N.G.C. Spooner and P.F. Smith, Phys. Lett. B 314 (1993) 430. \\
N.G.C. Spooner et al., Phys. Lett. B 321 (1994) 156.
\bibitem{sar107} M.L. Sarsa et al., Nucl. Phys. B (Proc. Suppl.) 48 (1996) 73;
Phys. Lett. B 386 (1996) 458; Phys. Rev. D56 (1997) 1856.
\bibitem{fus108} K. Fushimi et al., Phys. Rev. C 47 (1993) R425. \\
H. Ejiri et al., Phys. Lett. B 317 (1993) 14.
\bibitem{bac109} C. Bacci et al., Astropart. Phys. 2 (1994) 117.
\bibitem{bel110} P. Belli et al., Nuovo Cim. C 19 (1996) 537;
Nucl. Phys. B (Proc. Suppl.) 48 (1996) 62.
\bibitem{dru111} A. Drukier et al., Phys. Rev. D 33 (1986) 3495.
\bibitem{col112} J. Collar and F.T. Avignone, Phys. Rev. D47 (1993).
\bibitem{spe113} D.N. Spergel, Phys. Rev. D 37 (1988) 1353.
\bibitem{boo114} N. Booth, B. Cabrera and E. Fiorini, Ann. Rev. of Nucl. Sci.
46 (1996) 471.
\bibitem{cry115} in Cryogenic Detectors, LTD-5 Workshop (Berkeley 1993) (Journ of
Low Temp. Phys. 93, 1993); in LTD-7 Workshop (Garching, July 1997), ed.
S. Cooper (Pub. Max Planck Institut, Munich 1997).
\bibitem{sup116} Proc. Workshop on Superconductivity and Particle Detection
(Toledo 1994), eds. T. Girard, A. Morales and G. Waysand (Editions Fronti\`{e}res,
1995).
\bibitem{fio117} E. Fiorini, Nucl. Phys. B (Proc. Suppl.) 48 (1996) 41;
J. Low Temp. Phys. 93 (1993) 189.
\bibitem{lew118} J. D. Lewin, P.F. Smith, Astrop. Phys. 6 (1996) 87.
\bibitem{eng119} J. Engel, Phys. Lett. B264 (1991) 114. \\
J. Engel, S. Pittel and P. Vogel, J. Int. Mod. Phys. E 1 (1992) 1.
\bibitem{eng120} J. Engel et al., Phys. Rev. C 52 (1995) 2216. \\
J. Engel and P. Vogel, Phys. Rev. D40 (1989) 3132.
\bibitem{res121} M.T. Ressell and D.J. Dean, Preprint hep-ph/9702290
(Feb. 1997). \\
M.T. Ressell et al., Phys. Rev. D 48 (1993) 5519.
\bibitem{cal122} J.A. R. Caldwell and J.P. Ostriker, Ap.J. 251 (1981) 61. \\
J.N. Bahcall et al., Ap.J. 265 (1983) 730. \\
M. Turner, E. Gates and G. Gynk, Preprint astro-ph/9601168. \\
E.I. Gates, G. Gyuk and M.S. Turner, Ap. J. 449 (1995) L123; Phys. Rev. Lett.
74 (1995) 3724.
\bibitem{ell123} J. Ellis and R.A. Flores, Phys. Lett. B263 (1991) 259.
\bibitem{sar124} M.L. Sarsa et al., Nucl Phys. B (Proc. Suppl.)
35 (1994) 154; in Proc. Workshop Dark Side of the Universe (Rome, June 1993),
ed. R. Bernabei (World Scientific, Singapore, 1994) p. 216.
\bibitem{bel125} P. Belli et al. Nucl. Phys. B (Proc. Suppl.) 48 (1996) 61.\\
R. Bernabei et al., Phys. Lett. B389 (1996) 757.
\bibitem{spo126} N.G.C. Spooner (UKDMC) Nucl. Phys. B (Proc. Suppl.) 48
(1996) 64. \\
P.F. Smith et al., Phys. Lett. B379 (1996) 299.
\bibitem{bot127} A. Bottino et al., Phys. Lett. B 402 (1997) 113.
\bibitem{fre128} K. Freese et al., Phys. Rev. D 37 (1988) 3388.
\bibitem{bel129} R. Bernabei et al., Phys. Lett. B 424 (1998) 195.
\bibitem{ger130} G. Gerbier et al., Preprint astro-ph/9710181 (Oct. 1997).
\bibitem{bot131} A. Bottino et al., Phys. Lett. B 423 (1998) 109.
\bibitem{haz132} R. Hazama et al., in Proc. WEIN 95, eds. H. Ejiri et al.
(World Scientific, Singapore, 1995) p. 635; \\
H. Ejiri, in Proc. Conference on Neutrino Physics and Astrophysics, Neutrino 96
(Helsinki, June 1996), eds. K. Enqvist et al. (World Scientific, Singapore, 1997) p.342.
\bibitem{sum133} T.J. Sumner, in Proc. Workshop on Topics in Astroparticle and
Underground Physics, TAUP 97 (Gran Sasso, September 1997), eds. A. Bottino et al.,
to appear in  Nucl. Phys. (Proc. Suppl.) 1998.
\bibitem{ben134} P. Benetti et al., Nucl. Instrum. Methods Phys. Res. A 237 (1993) 203.
\bibitem{par135} J. Park et al., in Proc. Workshop on Sources of Dark Matter
in the Universe, ed. D. Cline (World Scientific, Singapore, 1994) p. 288.
\bibitem{zac136} V. Zacek, N. Cim. 107 A (1994) 291.
\bibitem{col137} J. Collar, Phys. Rev. D 54 (1996) 1247.
\bibitem{col138} J. Collar et al, in Proc. Workshop on the Identification of
Dark Matter (Sheffield, September 1996), ed. N.J. Spooner (World
Scientific, Singapore 1997) p. 563.
\bibitem{ham139} L.A. Hamel et al., in Proc. Workshop on the Identification of
Dark Matter (Sheffield, September 1996), ed. N.J. Spooner (World Scientific,
Singapore, 1997) p. 569. \\
V. Zacek, in Proc. Workshop on the Dark Side of the Universe (Rome, November 1995),
ed. R. Bernabei (World Scientific, Singapore, 1996).
\bibitem{ger140} G. Gerbier et al., Nucl. Phys. B (Proc. Suppl.) 13 (1990) 207.\\
K.N. Buckland et al., Phys. Rev. Lett. 73 (1994) 1067.
\bibitem{bro141} C. Broggini et al., Nucl. Phys. B (Proc. Suppl.) 35 (1994) 441.
\bibitem{ric142} J. Rich and M. Spiro, Sacaly Report DPhPE88-04. \\
C. Broggini and J.L. Vuilleumier, LNGS Report 92-1.
\bibitem{bel143} P. Belli et al., Nuovo Cimento 15C (1992) 475.
\bibitem{col144} J. Collar and F.T. Avignone, Astrop. Phys. 3 (1995) 37.
\bibitem{mar145} C.J. Martoff et al., Phys. Rev. Lett. 76 (1996) 4882.
\bibitem{ada146} J.S. Adams et al., in Proc. Workshop on the Identification of Dark
Matter (Sheffield, September 1996), ed. N.J. Spooner (World Scientific, Singapore,
1997) p. 469.
\bibitem{sno147} D. Snowden-Ifft et al., Phys. Rev. Lett. 70 (1993) 2348;
Phys. Rev. Lett.  74 (1994) 4133; Phys. Rev. Lett.  76 (1996) 331.
\bibitem{bau148} L. Baudis et al. (HMDM Coll.), in Proc. Workshop on Topics in
Astroparticle and Underground Physics, TAUP 97 (Gran Sasso, September 1997),
eds. A. Bottino et al., to appear in in Nucl. Phys. B (Proc. Suppl.) (1998).
\bibitem{tau150} Proc. Workshop on Topics in Astroparticle and Underground Physics,
TAUP Session 1995, eds. A. Morales et al., Nucl. Phys. B (Proc. Suppl.) 48 (1996);
and Session 1997, eds. A. Bottino et al., to appear in Nucl. Phys. B (Proc. Suppl.)
(1998).
\bibitem{she151} Proc. Workshop on The Identification of Dark Matter (Sheffield,
September 1996), ed. N.J. Spooner (World Scientific, Singapore, 1997).
\bibitem{sad152} B. Sadoulet, J. Low Temp. Phys. 93 (1993) 821.
\bibitem{fio153} E. Fiorini and T. Niinikoski, Nucl. Instrum. Methods Phys.
Res. 224 (1984) 83.
\bibitem{pre154} K. Pretzl, J. Low Temp. Phys. 93 (1993) 439; Particle World
1/3 (1990) 153. \\
B. Turrell, in Proc. Workshop on Low Temperature Detectors, LTD-7, (Garching 1997),
ed. S. Cooper (Pub. Max Planck Institute, Munich 1997).
\bibitem{way155} G. Waysand, G. Charding, eds. Superconducting and Low Temperature
Particle Detectors, Elsevier North Holland, 1989.
\bibitem{ale156} A. Alessandrello et al., Nucl. Phys. B (Proc. Suppl.)
35 (1994) 366; Nucl. Phys. B (Proc. Suppl.) 28 (1992) 233.
\bibitem{ale157} A. Alessandrello et al., Nucl. Instrum. Methods Phys. Res. A
360 (1995) 363. \\
E. Fiorini, Nucl. Phys. B (Proc. Suppl.) 48 (1996) 41.
\bibitem{ale158} A. Alessandrello et al., in Proc. Workshop on Low Temperature Detectors, LTD-7, (Garching 1997),
ed. S. Cooper (Pub. Max Planck Institute, Munich 1997) p. 246.
\bibitem{fio159} E. Fiorini et al., ``CUORE: A Cryogenic Underground Observatory for
Rare Events''. Letter of intent to the Gran Sasso Underground Laboratory.
December 1997.
\bibitem{cor160} N. Coron et al., Nucl. Phys. B (Proc. Suppl.) 35 (1994) 169.
\bibitem{yvo161} D. Yvon, Nucl. Phys. B (Proc. Suppl.) 48 (1996) 78.
\bibitem{sei162} W. Seidel et al., J. Low Temp. Phys. 93 (1993) 797. \\
P. Colling et al., Nucl. Instrum. Methods Phys. Res. A 354 (1995) 408.
\bibitem{sis163} M. Sisti et al., in Proc. Workshop on Low Temperature Detectors,
LTD-7, (Garching 1997), ed. S. Cooper (Pub. Max Planck Institute, Munich 1997) p. 232.\\
L. Zerle et al., in Proc. Workshop on Topics in
Astroparticle and Underground Physics, TAUP 97 (Gran Sasso, September 1997),
eds. A. Bottino et al., to appear in in Nucl. Phys. B (Proc. Suppl.) (1998).
\bibitem{bob164} C. Bobin et al, in Proc. Workshop on Topics in
Astroparticle and Underground Physics, TAUP 97 (Gran Sasso, September 1997),
eds. A. Bottino et al., to appear in in Nucl. Phys. B (Proc. Suppl.) (1998).
\bibitem{nam165} S.W. Nam et al., in Proc. Workshop on Low Temperature Detectors, LTD-7,
(Garching 1997), ed. S. Cooper (Pub. Max Planck Institute, Munich 1997), p. 217. \\
D.S. Akerib et al., in Proc. Workshop on Topics in
Astroparticle and Underground Physics, TAUP 97 (Gran Sasso, September 1997),
eds. A. Bottino et al., to appear in in Nucl. Phys. B (Proc. Suppl.) (1998).
\bibitem{hot166} D.L. Hote et al., in Proc. Workshop on Low Temperature Detectors, LTD-7,
(Garching 1997), ed. S. Cooper (Pub. Max Planck Institute, Munich 1997) p. 237. \\
Ph. Di Stefano, in Proc. Workshop on Topics in
Astroparticle and Underground Physics, TAUP 97 (Gran Sasso, September 1997),
eds. A. Bottino et al., to appear in in Nucl. Phys. B (Proc. Suppl.) (1998).
\bibitem{shu167} T. Shutt et al., Phys. Rev. Lett. 29 (1992) 3425. \\
P.D. Barnes et al., J. Low Temp. Phys. 93 (1993) 79.
\bibitem{gai168} R.J. Gaitskell et al., in Proc. Workshop on Low Temperature Detectors,
LTD-7, (Garching 1997), ed. S. Cooper (Pub. Max Planck Institute, Munich 1997) p. 221.
\bibitem{dav169} A.K. Davies et al, in Proc. Workshop on Low Temperature Detectors, LTD-7,
(Garching 1997), ed. S. Cooper (Pub. Max Planck Institute, Munich 1997) p. 227. \\
R.M. Clarke et al., in Proc. Workshop on Low Temperature Detectors, LTD-7,
(Garching 1997), ed. S. Cooper (Pub. Max Planck Institute, Munich 1997) p. 229.
\bibitem{abp170} M. Abplanalp et al., Nucl. Instrum. Methods Phys. Res. A 370 (1996) 227. \\
B. van den Brandt et al, in Proc. Workshop on Low Temperature Detectors, LTD-7,
(Garching 1997), ed. S. Cooper (Pub. Max Planck Institute, Munich 1997) p. 193. \\
K. Pretzl et al., in Proc. Workshop on Topics in Astroparticle and Underground
Physics, TAUP 97 (Gran Sasso, September 1997), eds. A. Bottino et al.,
to appear in in Nucl. Phys. B (Proc. Suppl.) (1998).
\bibitem{gir171} T.A. Girard et al., Nucl. Instrum. Methods Phys. Res. A 370 (1996) 223.
\bibitem{ber172} H. Bernas et al., Phys. Lett. A 24 (1967) 721. \\
A. Drukier and C. Valette, Nucl. Instrum. Methods Phys. Res. 105 (1972) 285.
\bibitem{boo173} N. Booth, Ap. Phys. Lett. 50 (1987) 293. \\
D.J. Goldie et al., Phys. Rev. Lett. 64 (1990) 954.
\end{thebibliography}
\end{document}